\documentclass[%
 reprint,
 amsmath,amssymb,
 aps,
]{revtex4-2}

\usepackage{graphicx}
\usepackage{dcolumn}
\usepackage{bm}

\makeatletter
\def\fourvdots{\vbox{\baselineskip1\p@ \lineskiplimit\z@\kern6\p@\hbox{.}\hbox{.}\hbox{.}\hbox{.}}}
\makeatother

\usepackage{physics}
\usepackage{bbm}
\usepackage[dvipsnames]{xcolor}

\begin{document}

\preprint{APS/123-QED}

\title{Microscopic imprints of learned solutions in adaptive resistor networks}

\author{Marcelo Guzman$^{1}$}
 \thanks{These two authors contributed equally}
\author{Felipe Martins$^{1}$}%
\thanks{These two authors contributed equally}
\author{Menachem Stern$^{2}$}
\author{Andrea J. Liu$^{1}$}
\affiliation{$^1$Department of Physics and Astronomy, University of Pennsylvania
}%

\affiliation{$^2$AMOLF, Science Park 104, 1098 XG Amsterdam, The Netherlands}

\date{\today}

\begin{abstract}
In physical networks trained using supervised learning, physical parameters are adjusted to produce desired responses to inputs.
An example is electrical contrastive local learning networks of nodes connected by edges that are resistors that adjust their conductances during training.
When an edge conductance changes, it upsets the current balance of every node. 
In response, physics adjusts the node voltages to minimize the dissipated power. 
Learning in these systems is therefore a coupled double-optimization process, in which the network descends both a \emph{cost landscape} in the high-dimensional space of edge conductances, and a \emph{physical landscape}--the power-- in the high-dimensional space of node voltages. Because of this coupling, the physical landscape of a trained network contains information about the learned task. Here we demonstrate that {\it all} the physical information relevant to the trained input-output relation can be captured by a susceptibility, an experimentally measurable quantity. We supplement our theoretical results with 
simulations to show that the susceptibility is positively correlated with functional importance and that we can extract physical insight into how the system performs the task from the conductances of highly susceptible edges.

\end{abstract}

\maketitle

\section{\label{sec:level1}Introduction}
In artificial neural networks, the cost landscape describes the value of the cost function, $C$, in terms of the parameters, or ``adaptable degrees of freedom" $k_i$~\cite{mehta2019high}.
Learned solutions correspond to low minima of this landscape, which are locally described by the cost Hessian~\cite{sagun2016eigenvalues,sagun2017empirical,dauphin2014identifying}, the matrix of second derivatives of the cost with respect to the adaptable degrees of freedom, $\mathcal H_{ij}=\partial_{k_i}\partial_{k_j}C$.  In the directions of the highest eigenmodes of the cost Hessian, corresponding to the highest positive curvatures, small changes in the adaptable degrees of freedom cause a substantial increase of the cost.  These key adaptable degrees of freedom reveal how the system achieves the task; for example, in classification they correspond to the decision boundary~\cite{sabanayagam2023unveiling,sagun2016eigenvalues}. 

In physical networks, physical parameters such as the stiffnesses or presence of springs connecting nodes in mechanical networks~\cite{rocks2017designing,pinson2017self,hexner2019effect,pashine2021local,arinze2023learning}, or conductances in resistor networks~\cite{stern2021supervised} can be tuned to achieve a desired physical response.
In particular, contrastive local learning networks~\cite{dillavou2022demonstration,dillavou2024machine}, use a local rule~\cite{stern2022learning} rather than gradient descent to adjust their conductances to reach minima in their cost landscapes, enabling them to perform supervised learning without using a processor. 
Like artificial neural networks, trained physical networks have directions of high curvature in the cost landscape~\cite{stern2024hessians} that identify the key adaptable resistors, or \emph{key edges}, responsible for performing the tasks. 

Unlike artificial neural networks, however, these physical networks are also described by a \emph{physical landscape}. For electrical contrastive local learning networks, this is the dissipated power $\mathcal P$ as a function of all the node voltages, or ``physical degrees of freedom" $V_a$.
The response to applied voltages or currents is defined by the minimum of $\mathcal P$, which translates to Kirchhoff's current law~\cite{vadlamani2020physics}.
The physical landscape for such systems is parabolic and thus fully described by the \emph{physical Hessian}, the second derivative of the power with respect to the node voltages, $\mathbf H_{ab}=\partial_{V_a}\partial_{V_b} P$.

Learning in electrical resistor networks therefore requires two optimizations.
On one hand, the system naturally responds to applied currents/voltages by settling into a voltage configuration that minimizes the power defined by the edge conductances. On the other hand, the system adjusts the edge conductances, modifying the power landscape and response, to minimize the cost function.
This double optimization leaves strong signatures in the physical modes for generic Hopfield~\cite{stern2024physical}, elastic~\cite{tlusty2017physical,yan2017architecture,yan2018principles,husain2020physical,stern2024physical}, and flow/resistor networks~\cite{anisetti2023emergent,stern2024physical}. These signatures include low-dimensional physical responses and alignment of low-lying eigenmodes  with the functional response~\cite{stern2024physical}. Most notably, stiff eigenmodes of the cost Hessian are related to soft eigenmodes of the physical Hessian~\cite{stern2024hessians}: for a task in which an applied voltage drop across one edge leads to an equal voltage drop across another edge, the lowest physical mode and the highest cost mode contain the same information~\cite{stern2024hessians}.
In this case, the lowest eigenmode of the physical Hessian $\mathbf H$ tells us the \emph{key edges}, namely the edges that are most responsible for performing the task. For more complex tasks, however, the situation is not as simple. Here we show precisely how much information about the cost Hessian is contained in the physical Hessian for arbitrarily complex tasks.

Our results provide a general way of identifying the key edges from physical properties of the networks alone. We demonstrate how this identification can be used to gain physical insight into the inner workings of trained networks--something that is not possible in artificial neural networks. 

\begin{figure}
\centering
\includegraphics[width=\columnwidth]{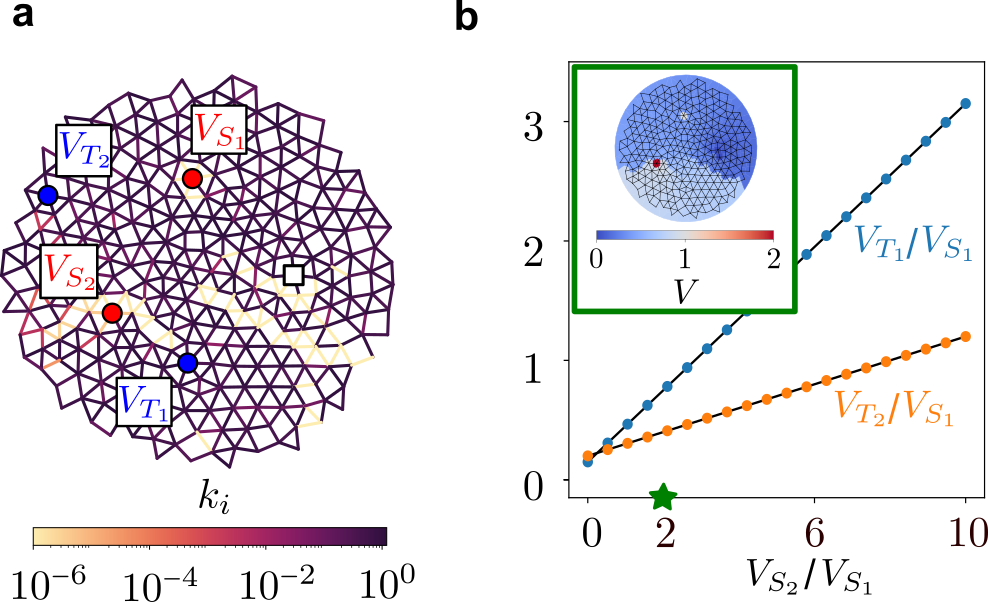}
\caption{\textbf{Linear regression task in a resistor network.}
{\bf a.} Conductance configuration satisfying a linear regression task between its input (red) and output (blue) nodes. The network is grounded at the white node, fixing the zero voltage reference.
{\bf b.} Target node voltages as a function of the source node voltages. To represent the relation in two dimensions, all variables are scaled by the first source voltage $V_{S_1}$. Circles correspond to the response of the trained circuit and solid lines to the desired relation. 
Inset: Voltage response (color) when applying the inputs indicated by the green star: $V_{S_1} = 1$ and $V_{S_2} = 2$.
}
\label{fig:example}
\end{figure}
\section{Systems Studied}
We will couch our analysis in the language of adaptable resistor networks of $N$ nodes and $M$ edges, in which each edge corresponds to a linear resistor with adjustable conductance.
Each node $a$ in the network is characterized by a voltage $V_a$ with respect to a ground. Each edge $i$ connecting two nodes $a$ and $b$ carries a current proportional to the voltage difference, $I_{i}=k_{i}(V_a-V_b)$, with $k_{i}$ being the conductance of the edge connecting nodes $a$ and $b$.
Throughout the article we reserve indices $\{a,b\}$ to indicate nodes and $\{i,j\}$ to indicate edges. We use vectors to describe voltage states $\mathbf V=(V_1,V_2,...,V_N)$, conductance states $\bm k =(k_1,k_2,...,k_M)$, and current states $\mathbf I =(I_1,I_2,...,I_M)$. 

For the numerical analyses, we use adaptable resistor networks of size $N=300$ nodes obtained from jammed packings, as in previous studies~\cite{rocks2019limits,stern2021supervised}.
Inspired by real laboratory implementations, we constrain edge conductances to lie within a finite but wide range of $[10^{-6},10^6]$.
We train the networks using Coupled Learning~\cite{stern2021supervised}, but emphasize that our theoretical results are independent of the training process. The main assumption is that the cost is near the global minimum value of zero, so the task has been learned.
For training details, initial conditions, and hyperparameters of the different examples shown in the paper, see Appendix~\ref{sec:A2}.

\section{Relation between the cost and physical Hessians}
The following derivation holds for any trained task. However, it is useful to have in mind an illustrative example. As such an example, consider a network trained for a linear regression task, fig.~\ref{fig:example}.
The network has two source nodes $S_1$ and $S_2$, and two target nodes $T_1$ and $T_2$, satisfying the following relation:
\begin{equation}
    \mathbf V^{\text{des}} \equiv\begin{pmatrix}
        V_{T_1}\\V_{T_2}
    \end{pmatrix} =\begin{pmatrix}
        0.15 & 0.3\\
        0.2 & 0.1
    \end{pmatrix}\begin{pmatrix}
        V_{S_1}\\
        V_{S_2}
    \end{pmatrix}.
    \label{eq:task}
\end{equation}

In addition, we hold a node at ground voltage (Fig.~\ref{fig:example}a), as in experiments.
This removes the trivial zero modes of the physical landscape corresponding to uniform shifts of voltage on all the nodes.
The physical Hessian for a grounded linear resistor network is given by
\begin{equation}
\mathbf H =\begin{pmatrix}2 \Delta K\Delta^T & \bm q\\\bm q^T&0\end{pmatrix},
\label{eq:physicalhessian1}
\end{equation}
where $\Delta$ is the incidence matrix, $K_{ij}=\delta_{ij}k_i$ the diagonal matrix of conductances, and $q_a=\delta_{a,a'}$ is a canonical vector with entry 1 at the fixed ground node $a'$.
This {\it extended} Hessian or bordered Laplacian~\cite{stern2024hessians,rocks2019limits} removes the trivial zero mode of the network, allowing the physical Hessian to be inverted, $\mathbf H^{-1}$.
Under this formulation, the voltage configuration is extended with the current (Lagrange multiplier) $\lambda$ needed to ground the system, $\mathbf V = (V_1,V_2,...,V_N,\lambda)$.

Additional external input currents associated e.g. with training samples, can be generically encoded into a current vector of the form $\mathbf I = \left(I_1, I_2, ..., I_N,0\right)$, where the last entry corresponds to the value corresponding to the ground voltage.
The system's voltage response $\mathbf V^F$ to all of the applied currents must minimize the dissipated power $\mathcal P$
\begin{equation}
\mathcal P = \frac{1}{2}\mathbf V^T\mathbf H\mathbf V - \mathbf I^T \mathbf V.
\label{eq:P1}
\end{equation}
leading to,
\begin{equation}
\mathbf V^F  = \mathbf H^{-1}\mathbf I.
\label{eq:freestate}
\end{equation}

\begin{figure}
\centering
\includegraphics[width=0.9\columnwidth]{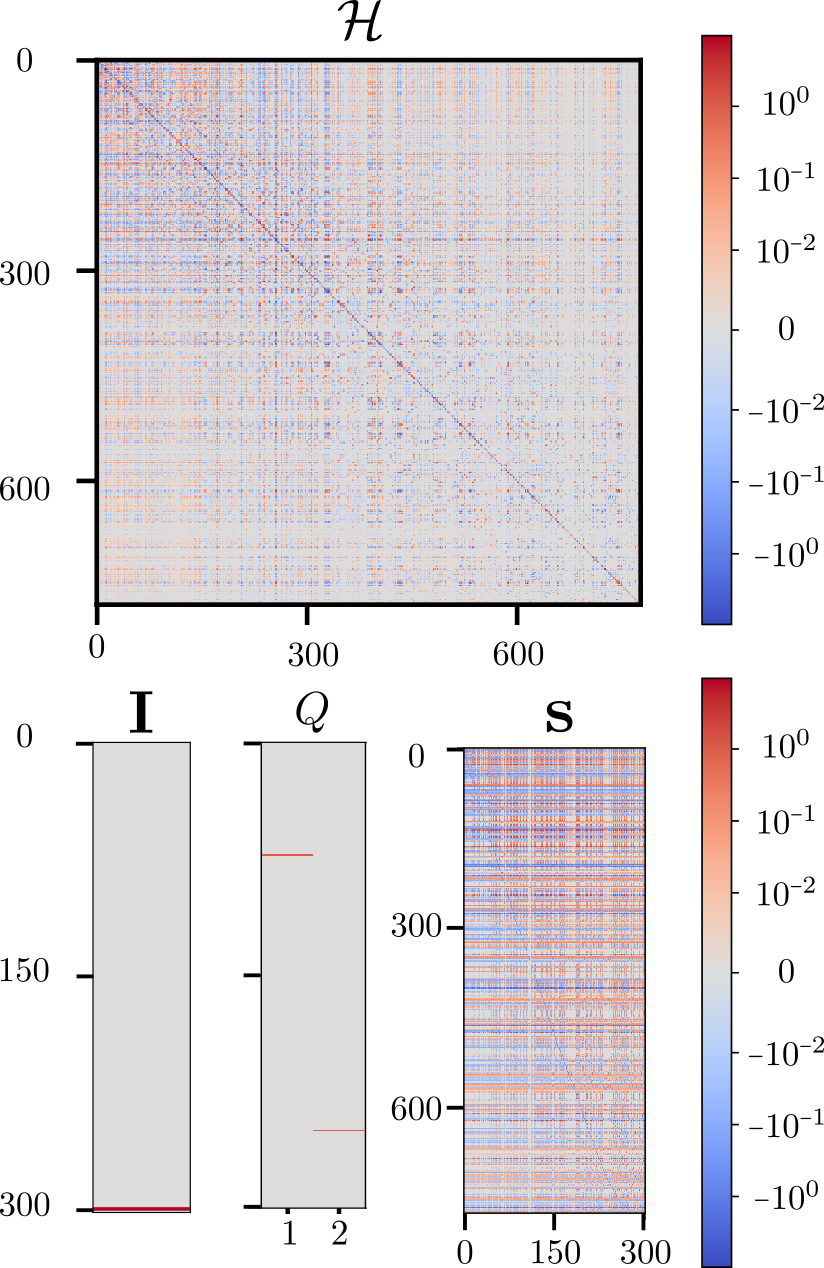}
\caption{\textbf{Learning and physical contributions to the cost Hessian.}
Entries of the cost Hessian $\mathcal H$, input current $\mathbf I$, output projector $Q$, and susceptibility matrix $\mathbf s$ for the trained network shown in fig.~\ref{fig:example}.
}
\label{fig:hessiandecomposition}
\end{figure}
Supervised learning corresponds to modifying the response $\mathbf V_r^F$ to a given set of inputs $\mathbf I_r$ to satisfy constraints $\bm c_r(\mathbf V_r^F)=\bm 0$.
Usually, these constraints act on specific nodes designated as output voltage nodes.
Denoting $Q_r$ the projector onto these output nodes for task $r$, each constraint can be written as
\begin{equation}
\bm c_r(\mathbf V_r^F) = Q_r \mathbf V_r^F-\mathbf V_r^{\text{des}},
\label{eq:constraint}
\end{equation}
where $\mathbf V_r^{\text{des}}$ is the vector of desired output values for the intputs $\mathbf I_r$.
Notice that the length of  $\bm c_r$ and $\mathbf V_r^{\text{des}}$ is equal to the number of output nodes.

The cost $C$ is then naturally defined as the sum of the squared constraints over the $R$ tasks the networks is being trained, corresponding to the mean squared error (MSE):
\begin{equation}
C=\frac{1}{2}\sum_{r=1}^{R} \bm c_r^T\bm c_r.
\label{eq:cost}
\end{equation}
Under the assumption that all the constraints are satisfied at the end of training, $\bm c_r =\bm 0,\;\;\; \forall r$, the cost Hessian reads~\cite{stern2024hessians}:
\begin{equation}
\mathcal H_{ij} = \frac{\partial ^2 C}{\partial k_i\partial k_j}= \sum_{r=1}^R \frac{\partial \bm c_{r}^T}{\partial k_i}\frac{\partial \bm c_{r}}{\partial k_j}.
\label{eq:costHessian}
\end{equation}
Using Eqs.~\eqref{eq:constraint} and \eqref{eq:freestate}, we carry out the derivatives of the constraints over the voltage response,
\begin{equation}
\frac{\partial\bm c_r}{\partial k_i} =Q_r \frac{\partial\mathbf V_r^F}{\partial k_i} = Q_r \frac{\partial\mathbf H ^{-1}}{\partial k_i}\mathbf I_r,
\end{equation}
leading to 
\begin{equation}
\mathcal H_{ij} = \sum_{r=1}^R \mathbf I_r^T \frac{\partial\mathbf H ^{-1}}{\partial k_i} Q_r^T Q_r \frac{\partial\mathbf H ^{-1}}{\partial k_j}\mathbf I_r.
\label{eq:hessian1}
\end{equation}
This expression highlights two contributions to the cost Hessian.
First, there is a physical contribution stemming from $\partial \mathbf H^{-1}/\partial k_i$. This contribution is agnostic to training details and therefore depends only on the conductance values and the network topology.
Second, there is a training contribution that depends on how the system is trained, namely the identities of the output nodes embodied in $Q_r$, and the input values for each task, $\mathbf I_r$.

We can explicitly separate physical and training contributions of the cost Hessian by employing higher-order tensors.
Denoting the outer product with $\otimes$, we define the 4th-rank training tensor $\mathbf L$ and the 4th-rank susceptibility tensor $\mathbf S_{ij}$ as
\begin{align}
\mathbf L &= \sum_{r=1}^R\mathbf I_r^T\otimes Q_r^TQ_r\otimes \mathbf I_r\\
\mathbf S_{ij} &= \frac{\partial\mathbf H ^{-1}}{\partial k_i} \otimes \frac{\partial\mathbf H ^{-1}}{\partial k_j},
\label{eq:transmittance}
\end{align}
from which we can express the cost Hessian as the full contraction of these last two tensors:
\begin{equation}
\mathcal H_{ij}= \mathbf L\fourvdots\mathbf S_{ij}=\sum_{a,b,c,d}\mathbf L_{abcd}\mathbf S_{ij,abcd}.
\label{eq:hessian2}
\end{equation}

Equation~\eqref{eq:hessian2} clearly establishes that the susceptibility tensor contains all the physical information of the cost Hessian.
Its components measure the sensitivity of the inverse of the physical Hessian (the response) to changes in conductances.
Notably, the entries of this tensor are modified by the training components $\mathbf L$, yet they do not depend explicitly on them.
We highlight that while the derivation of Eq.~\eqref{eq:hessian2} assumed the MSE as the cost function, eq.~\eqref{eq:cost}, this result is valid for any form of cost provided it is differentiable and has a minimum, see Appendix~\ref{sec:A1}.

In what follows we provide a simpler analytical expression of the susceptibility tensor and link it to the linear response of the system.

\section{The susceptibility tensor}
Using Eq.~\eqref{eq:physicalhessian1} and the derivative of the matrix inverse, we explicitly compute the derivatives of the inverse physical Hessian appearing in Eq.~\eqref{eq:transmittance}:
\begin{equation}
\frac{\partial \mathbf H^{-1}}{\partial k_i} = -\mathbf H^{-1}\frac{\partial \mathbf H}{\partial k_i}\mathbf H^{-1} = 
 -\mathbf H^{-1}\begin{pmatrix}
 2  (\Delta^T_i)^T \Delta^T_i & \bm 0\\\bm 0^T&0
 \end{pmatrix}\mathbf H^{-1},
\end{equation}
where $\Delta^T_i$ is vector formed by the $i$-th row of the incidence matrix $\Delta$.
For clarity we define the extended  vector (bold font) $\mathbf \Delta^T_i = (\Delta^T_i, 0)$ by adding an extra zero, and the susceptibility vector $\mathbf s_i$ as
\begin{equation}
\mathbf s_i \equiv \mathbf \Delta_i^T\mathbf H^{-1},
\label{eq:susceptibility}
\end{equation}
corresponding to a vector of dimensions $N+1$ defined per edge $i$.
Under this notation, the derivative of the inverse physical Hessian reads
\begin{equation}
\frac{\partial \mathbf H^{-1}}{\partial k_i} = -2\; \mathbf s_i\otimes \mathbf s_i,
\end{equation}
yielding the final expression for the susceptibility tensor:
\begin{equation}
\mathbf S_{ij} = 4\; \mathbf s_i\otimes \mathbf s_i\otimes\mathbf s_j\otimes \mathbf s_j.
\label{eq:susceptibilityTensor}
\end{equation}

At this point, the power dependence of both Hessians becomes evident.
Eq.~\eqref{eq:susceptibility} explicitly shows that the susceptibility vector scales as the inverse of the physical Hessian, $\mathbf s_i\sim \mathbf H^{-1}$.
Applying this scaling relation four times, as per eq.~\eqref{eq:susceptibilityTensor}, we obtain $\mathbf S_{ij}\sim \mathbf H^{-4}$, and ultimately $\mathcal H\sim \mathbf H^{-4}$.

While each entry $\mathbf S_{ij}$ is a 4th rank tensor, rendering $\mathbf S$ a 6th order tensor, its components are all determined by the susceptibility vectors $\mathbf s_i$ defined per edge.
Moreover, the susceptibility vectors are dense as opposed to the sparse nature of the input currents $\mathbf I_r$ and the projectors $Q_r$, fig.~\ref{fig:hessiandecomposition}.
Simply put, the larger the susceptibility $\mathbf s_i$, the larger the entry $\mathbf S_{ij}$, which, on average and depending on the training tensor $\mathbf L$, tends to manifest as large entries of the cost Hessian $\mathcal H_{ij}$.
We thus turn our attention to the susceptibility vectors and their norms.

\begin{figure}
\centering
\includegraphics[width=0.8\columnwidth]{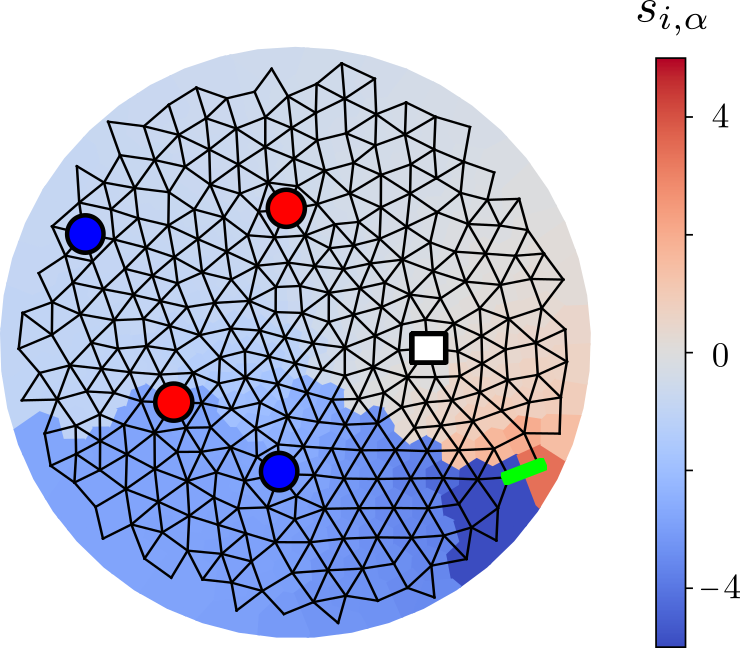}
\caption{\textbf{Susceptibility vector.}
Example of susceptibility vector $\mathbf s_i$ for the highlighted edge of the trained network of fig.~\ref{fig:example}.
In general, the closer the node $\alpha$ is to the edge, the larger the entry of $\mathbf s_{i,\alpha}$.
}
\label{fig:susceptibility}
\end{figure}

The definition of $\mathbf s_i$ is directly related to the response of the system. 
Contracting it with an input current $\mathbf I$, it leads to voltage drop at edge $i$ of the physical response to the current $\mathbf I$,
\begin{equation}
    \mathbf s_i^T \mathbf I = \mathbf \Delta _i^T\left(\mathbf H^{-1}\mathbf I \right).
\end{equation}
Thus each element $\mathbf s_{i,a}$ is the voltage drop at edge $i$ of the response to a unit current at node $a$, fig.~\ref{fig:susceptibility}.
The magnitude of $\mathbf s_i$  quantifies the voltage drop linked to all possible unit currents in the network and assigns a scalar measure of susceptibility per edge:
\begin{equation}
    ||\mathbf s_i||^2 = \sum_a||\mathbf s_{i,a}||^2.
\end{equation}

Denoting by $\nu_{\alpha}$ and $\mathbf W_{\alpha}$ the eigenvalues and normalized eigenvectors of the physical Hessian, $\mathbf H$, we can recast the susceptibility as
\begin{equation}
\mathbf s_i = \sum_{\alpha}\frac{1}{\nu_{\alpha}}\left(\mathbf \Delta_i^T\mathbf W_{\alpha}\right) \mathbf W_{\alpha}^T,
\end{equation}
with norm 
\begin{equation}
|| \mathbf s_i||^2 = \sum_{\alpha}\frac{1}{\nu_{\alpha}^2}\left(\mathbf \Delta_i^T \mathbf W_{\alpha}\right)^2,
\label{eq:norm}
\end{equation}
where we have used the orthonormality of the eigenvectors $\mathbf W_{\alpha}$.
Equation~\eqref{eq:norm} shows that the magnitude of the susceptibility is dominated by soft modes (small $\nu_\alpha$).
Thus, edges with large susceptibility must have large voltage drops across the modes, specially the softest.

In summary, we have demonstrated that all the physical information of the cost Hessian boils down to the dense susceptibility tensor $\mathbf S$ and ultimately to the susceptibility vectors $\mathbf s_i$.
The susceptibilities are dominated by soft modes of the physical Hessian with large voltage drops, and correlate positively with the cost Hessian. 

These results, although written in the language of resistor networks, are far more general.
There are only three necessary conditions to derive Eqs.~\eqref{eq:hessian2} and \eqref{eq:susceptibilityTensor}:
\begin{itemize}
    \item[I.] The adaptable degrees of freedom satisfy all the constraints and therefore $C=0$ (eq.~\eqref{eq:cost}). In other words, the system has been trained successfully.
    \item[II.] The physical response to externally applied inputs $\mathbf I$ minimizes a scalar function $\mathcal P$ with respect to the physical degrees of freedom, $\mathbf V^F=\min_{\mathbf V}\mathcal P(\mathbf V,\mathbf I)$. In other words, the physical system optimizes a Lyapunov function.
    \item[III.] The response $\mathbf V^F$ to inputs $\mathbf I$ is approximated by the perturbations around a known local minimum $\mathbf V_{\text{min}}$ of $\mathcal P$, $\mathbf V^F\approx \mathbf V_{\text{min}}+\delta \mathbf V$.
\end{itemize}
In low Reynolds number flow networks, the adaptable degrees of freedom may correspond to pipe diameters, the physical degrees of freedom to node pressures, and the scalar function $\mathcal P$ to the power dissipated.
In elastic networks, the adaptable degrees of freedom may taken to be the stiffness or rest lengths of all of the springs, the physical degrees of freedom are the node positions or displacements, and $\mathcal P$ is the elastic energy or free energy.

Conditions II and III allow for a generic perturbative description of nonlinear physical landscapes in terms of the closest local minimum $\mathbf V_{\text{min}}$ to the response $\mathbf V^F$.
In such case, eq.~\eqref{eq:P1} becomes $\mathcal P = P(\mathbf V)-\mathbf I^T\mathbf V$, with $P$ a nonlinear function of the physical degrees of freedom $\mathbf V$. To second order,
\begin{equation}
    \mathcal P(\mathbf V_{\text{min}}+\delta\mathbf V)\approx \mathcal P(\mathbf V_{\text{min}})+\frac{1}{2}\delta\mathbf V^T\mathbf H\delta\mathbf V- \mathbf I^T\delta\mathbf V,
\end{equation}
where the physical Hessian is given by $\mathbf H_{ab}=\partial _{V_a}\partial_{V_b}P(\mathbf V_{\text{min}})$.
Then, the response is given by $\mathbf V^F =\mathbf V_{\text{min}}+\mathbf H^{-1}\mathbf I$, and the same derivations ensue from eq.~\eqref{eq:constraint}.

In the remainder of the paper, we work exclusively with resistor networks to elucidate the physical significance of highly-susceptible edges.
Since they capture much of the information of the cost Hessian, we delve into how much they contribute to its stiff modes. Recall that stiff directions in the cost landscape correspond to directions in which the cost is particularly sensitive to changes of parameters (adaptable degrees of freedom). 

In the next sections we analyze three different learning tasks, previously implemented in experimental networks~\cite{dillavou2022demonstration,dillavou2024machine}, and demonstrate that the susceptibility is highly correlated with stiff modes of the cost Hessian, and therefore to the degree of sensitivity of the cost to the adaptable degrees of freedom corresponding to those directions in the cost landscape.
Finally, we will connect the  susceptibility to the low-dimensional response observed in trained networks~\cite{stern2024physical} and the topological nature of the response in allosteric networks~\cite{rocks2020StructureFunction}.

\section{Stiff modes and susceptible edges}

\begin{figure}
\centering
\includegraphics[width=\columnwidth]{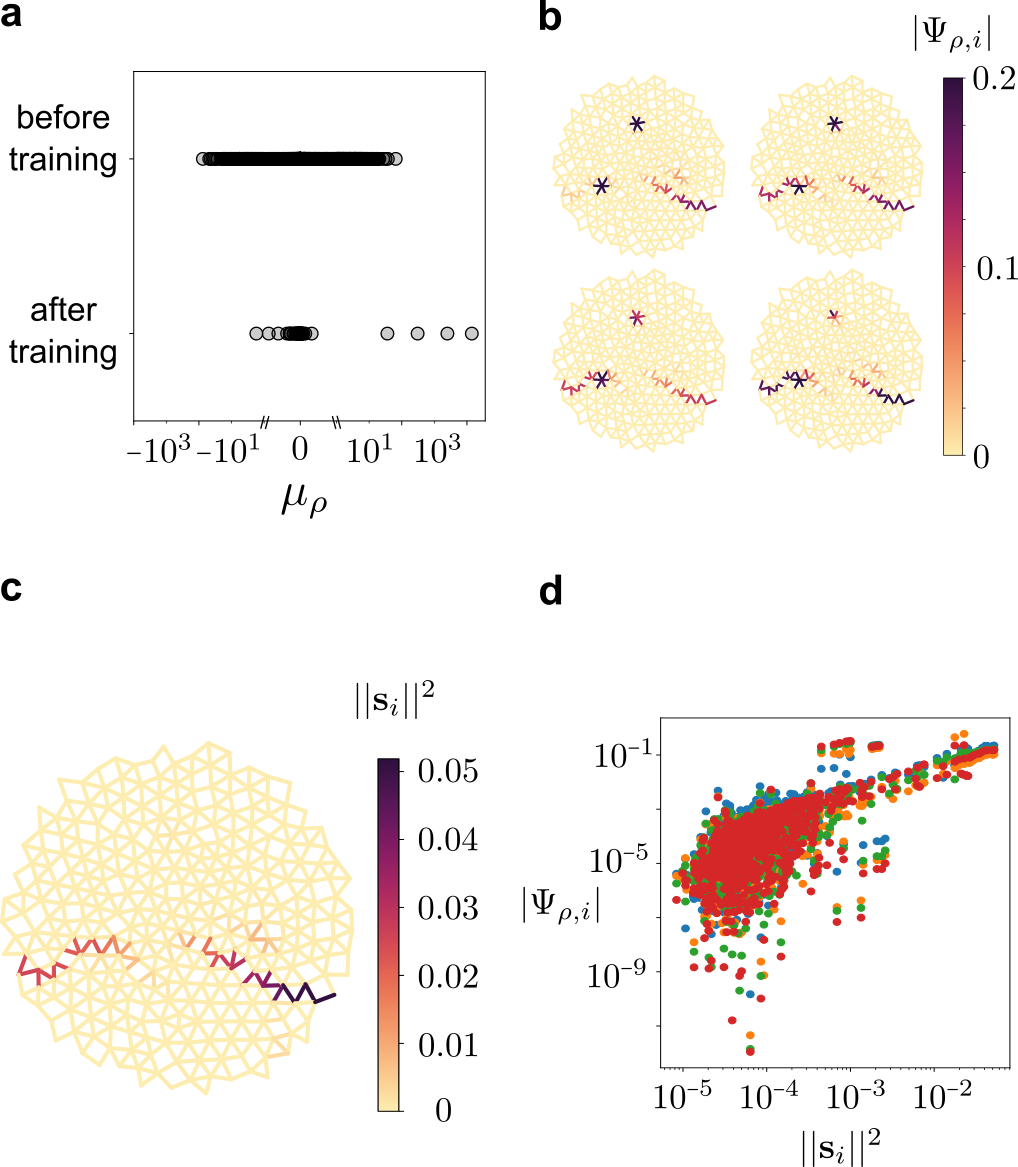}
\caption{\textbf{Stiff modes and susceptibility norm.}
{\bf a.} Eigenvalues of the cost Hessian, $\mu_{\rho}$, before and after training.
The spectrum has a spectral gap spanning more than 2 orders of magnitude, separating four stiff outliers from the remaining almost-zero modes.
{\bf b.} The four associated eigenvectors depicted as scalar fields over the networks. 
Dark colors highlight the conductances that are most important to the function.
{\bf c.} The scalar field $||\mathbf s_i||^2$ captures the main structure revealed by stiff modes of the cost Hessian in (b).
{\bf d.} Scatter plot of the absolute values of the entries of each of the stiff modes $|\Psi_{\rho,i}|$ and susceptibilities $||\mathbf s_i||^2$ for each edge in the network.
Different colors indicate different stiff modes.
}
\label{fig:stiff_suscept_regression}
\end{figure}

The description of the cost landscape near the solution is fully encoded in the cost Hessian $\mathcal H$.
Its eigenvalues, $\mu_{\rho}$, and eigenvectors $\Psi_{\rho}$, describe the curvature along different directions.
Figure~\ref{fig:stiff_suscept_regression}a shows the distribution of eigenvalues $\mu_{\rho}$ before and after training.
Before training, the network has a wide distribution of cost eigenvalues and a non-zero cost gradient.
After training, the spectrum is highly degenerate. Most of the eigenvalues are close to zero while four are very high. The same behavior has been observed in artificial neural networks~\cite{sagun2016eigenvalues,sagun2017empirical}.
The modes with high eigenvalues, or stiff modes, correspond to the number of coefficients learned by the linear regression task (details in fig.~\ref{fig:example}).
The entries of each of the stiff modes $\Psi_{\rho}$ measure the importance of the given edge to the function (Fig.~\ref{fig:stiff_suscept_regression}b). 
Perturbing any of those edges amounts to moving in conductance space along, at least partially, directions of high-curvature, substantially increasing the cost $C$.
Remarkably, the scalar field of the norm of susceptibilities shown in Fig.~\ref{fig:stiff_suscept_regression}c reveals the same patterns as we see in the stiff modes, Fig.~\ref{fig:stiff_suscept_regression}b.
In Fig.~\ref{fig:stiff_suscept_regression}d we quantify this observation with a correlation scatter plot over the edges of the network, showing that high-susceptibility correlate strongly with large entries of the stiff modes.

\begin{figure*}
\centering
    \includegraphics[width=\textwidth]{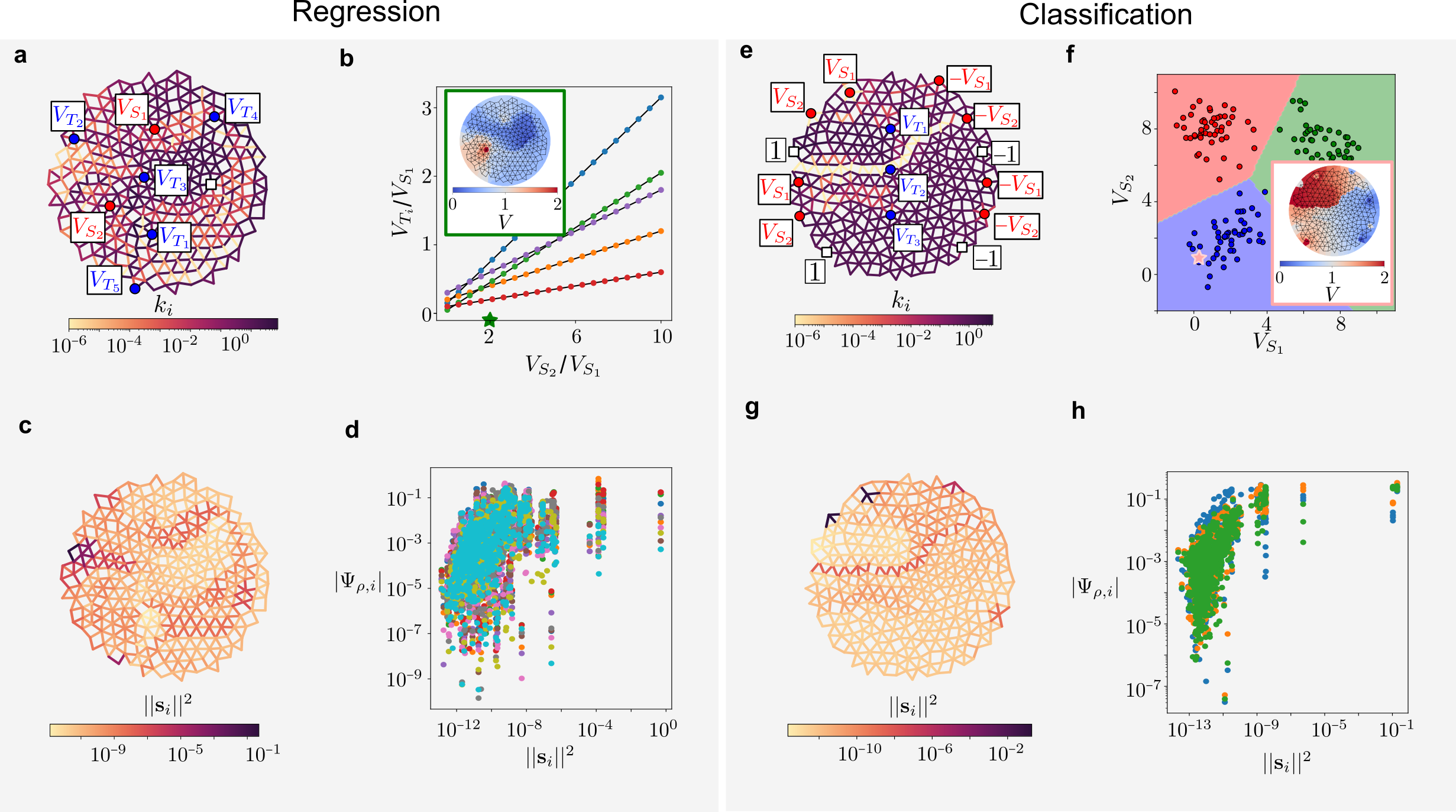}
\caption{\textbf{Stiff modes and susceptibility for linear regression and classification examples}\\
\textbf{Regression: }{\bf a.} Conductance configuration of a network trained for the five target linear regression task specified by the coefficients in ~\eqref{eq:5T_regression_coeffs}. Source nodes are denoted by red circles, target nodes by blue circles and the grounded node by a white square.
{\bf b.} The ratio of the voltage at each of the target nodes $V_{T_i}$ with respect to the voltage source $V_{S_1}$ (circles) obeys the required linear dependence (solid lines). Inset: voltage response associated to the green star ($V_{S_2}=2$ and $V_{S_1}=1$).
{\bf c.} Scalar field of $||\mathbf s_i||^2$, with darker colors indicating larger susceptibilities.
{\bf d.} The edges of higher susceptibility positively correlate with the entries of the 10 stiffest modes of the network.
\textbf{Classification: } Same panels as before for a network trained to classify three classes of data as shown in {\bf f}, where the inset corresponds to the the voltage response to the inputs indicated by the pink star.
}
\label{fig:stiff_suscept_regression2_class}
\end{figure*}

\begin{figure*}[t]
\centering
\includegraphics[width=\textwidth]{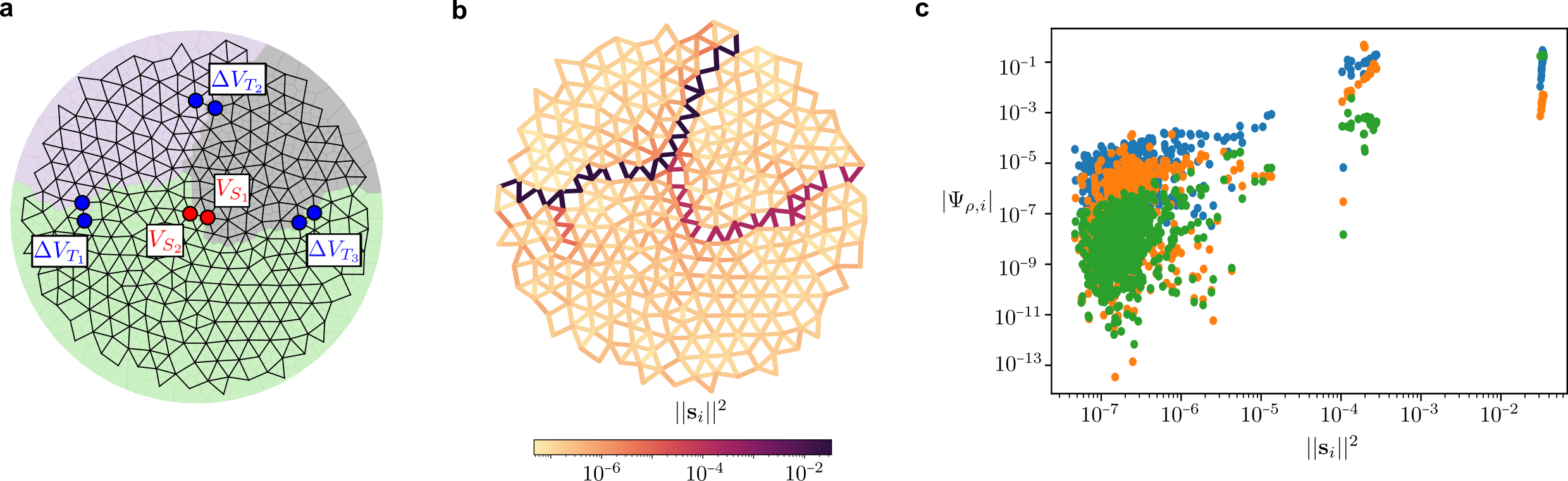}
\caption{\textbf{Susceptible edges define topological sectors.}
{\bf a.} Network trained for an allosteric task in which pairs of target nodes (blue) must have a voltage drop of 0.5, when the voltages at the input nodes (red) are 0 and 1.
The background color corresponds to the different topological sectors found by the persistent homology analysis of the voltage response (for details see~\cite{rocks2021hidden}).
{\bf b.} Without the information of input and output nodes and values, the susceptibility field explicitly captures the boundaries between the persistent sectors.
{\bf c.} For large values, the susceptibility positively correlates with the entries of the three stiffest modes (different colors).
}
\label{fig:edge_allo}
\end{figure*}

We illustrate the strong correlation between edge susceptibility and stiff modes of the cost Hessian for three more tasks.
First we consider a more complicated linear regression task, consisting of two input and 5 output nodes.
The vector of desired voltages as a function of input voltages is given by
\begin{equation}
    \mathbf V^{\text{des}} = \begin{pmatrix}
        0.15 & 0.3\\
        0.2 & 0.1\\
        0.05 & 0.2\\
        0.1 & 0.05\\
        0.3 & 0.15
    \end{pmatrix}\begin{pmatrix}
        V_{S_1}\\
        V_{S_2}
    \end{pmatrix},
    \label{eq:5T_regression_coeffs}
\end{equation}
where as before, a ground node is kept at zero voltage, making all the linear relations non-colinear.

The network is trained down to a cost of $C\approx 10^{-9}$, ending up with a set of conductances that satisfies the task, Fig.~\ref{fig:stiff_suscept_regression2_class}a and b.
As with the previous linear regression task, high susceptibility edges  positively correlate with those singled out from stiff modes, fig.~\ref{fig:stiff_suscept_regression2_class}c and d, 

A more stringent test corresponds to linear classification.
Here we train the network to classify three different classes depending on two input voltages, fig.~\ref{fig:stiff_suscept_regression2_class}e and f.
We encode the corresponding classes with three output nodes. To convert classification into a numerical task, we use ``one-hot encoding" for each output node.
For example, a point belonging to class 2, corresponds to $\mathbf V^{\text{des}} = (0, 1, 0)$.
For this case, a more suitable cost function corresponds to the cosine similarity:
\begin{equation}
   \mathcal L = \sum_r\left(1-\cos(\mathbf V^F_r,\mathbf V^{\text{des}})\right),
\end{equation}
where $\cos(\bm V_T,\bm V_T^{\text{des}})$ corresponds to the cosine of the angle between the two vectors.

This is a significantly harder task for the network to achieve for three reasons.
First, the physical restriction of having positive conductances only allows for decision boundaries with positive slope. To overcome this, we represent each input value as two different and opposite voltage nodes, Fig.~\ref{fig:stiff_suscept_regression2_class}e.
We do the same for the ground nodes, placing one low (-1) and high (+1) ground.
Second, the two-dimensional embedding of the network topologically constraints current flows, leading to output nodes being effectively isolated from input nodes.
We solve this by doubling the information and copying each input node twice, as well as the low and high grounds.
Third, achieving perfect one-hot encoding is impossible for a linear resistor network, since it corresponds to a highly non-linear output as a function of the input voltages. 
Nevertheless, we can still train and interpret the result using a winner-take-all strategy, in which the network classifies according to the maximum target value, $\max(V_{T_1},V_{T_2},V_{T_3})$. 
Figure.~\ref{fig:stiff_suscept_regression2_class}f shows that the network trains successfully, finding a set of edge conductances that leads to a classification accuracy of $99\%$.
Even though the training and task are more involved, the susceptibility vector is a good proxy of the stiff modes, as shown in fig.~\ref{fig:stiff_suscept_regression2_class}g and h.

\subsection{Connection to persistent homology}
As a last family of examples, we analyze networks trained to provide specified voltage drops across output edges in response to a voltage drop across an input edge~\cite{rocks2019limits} (Fig.~\ref{fig:edge_allo}a).
Such long-range effects require precise values of the conductances in resistor or flow networks~\cite{rocks2019limits}.
Compared to the previous regression and classification tasks, this task is significantly simpler due to the fixed input values it is trained on.
One of the remarkable results for networks trained for such tasks is that the response is topological in structure~\cite{rocks2020StructureFunction,rocks2021hidden, rocks2024topological}.
Regardless of the values the networks is trained on, the response of the system tends to partition into sectors of roughly homogeneous voltages, captured by a topological data analysis approach known as persistent homology, as shown in Fig.~\ref{fig:edge_allo}a~\cite{rocks2021hidden,rocks2020StructureFunction}.
It turns out that the boundaries of the topological sectors are precisely the edges with large susceptibility (Fig.~\ref{fig:edge_allo}b). These also correspond to edges singled out by stiff modes of the cost Hessian (Fig.~\ref{fig:edge_allo}c).
In Appendix \ref{sec:PH}, we show that the susceptibility is even better at capturing the important edges for weak training signals.

\section{Low-dimensional response}
\begin{figure}
\centering
\includegraphics[width=\columnwidth]{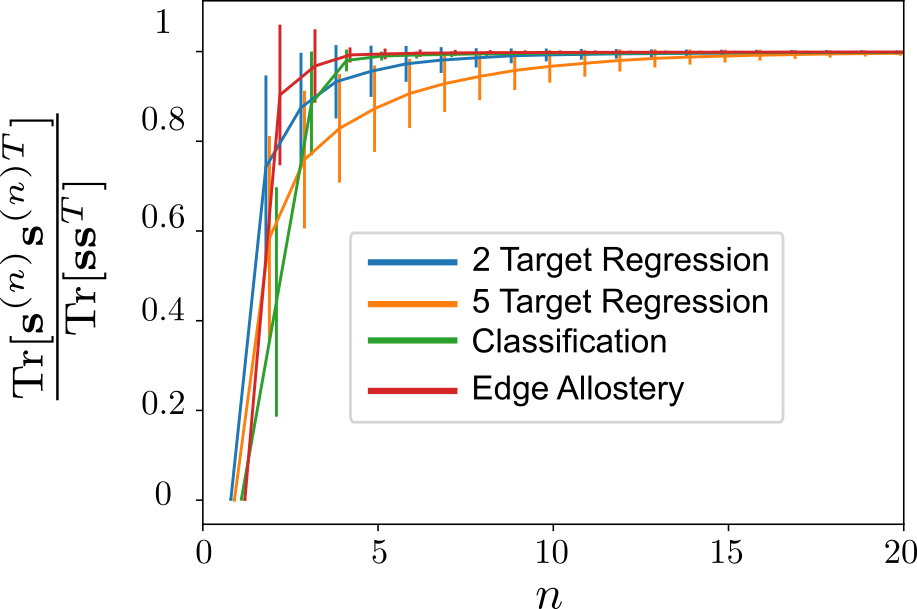}
\caption{\textbf{Susceptibility and dependence on number of modes.}
Ratio of partial to full susceptibility as a function of number of modes for the 4 kinds of tasks treated in the main text.
Error bars correspond to standard deviations over ensembles of 50 different initial conditions for each case.
The curves are slightly shifted horizontally with respect to each other to distinguish the different error bars.
}
\label{fig:size}
\end{figure}
\begin{figure*}
\centering
\includegraphics[width=\textwidth]{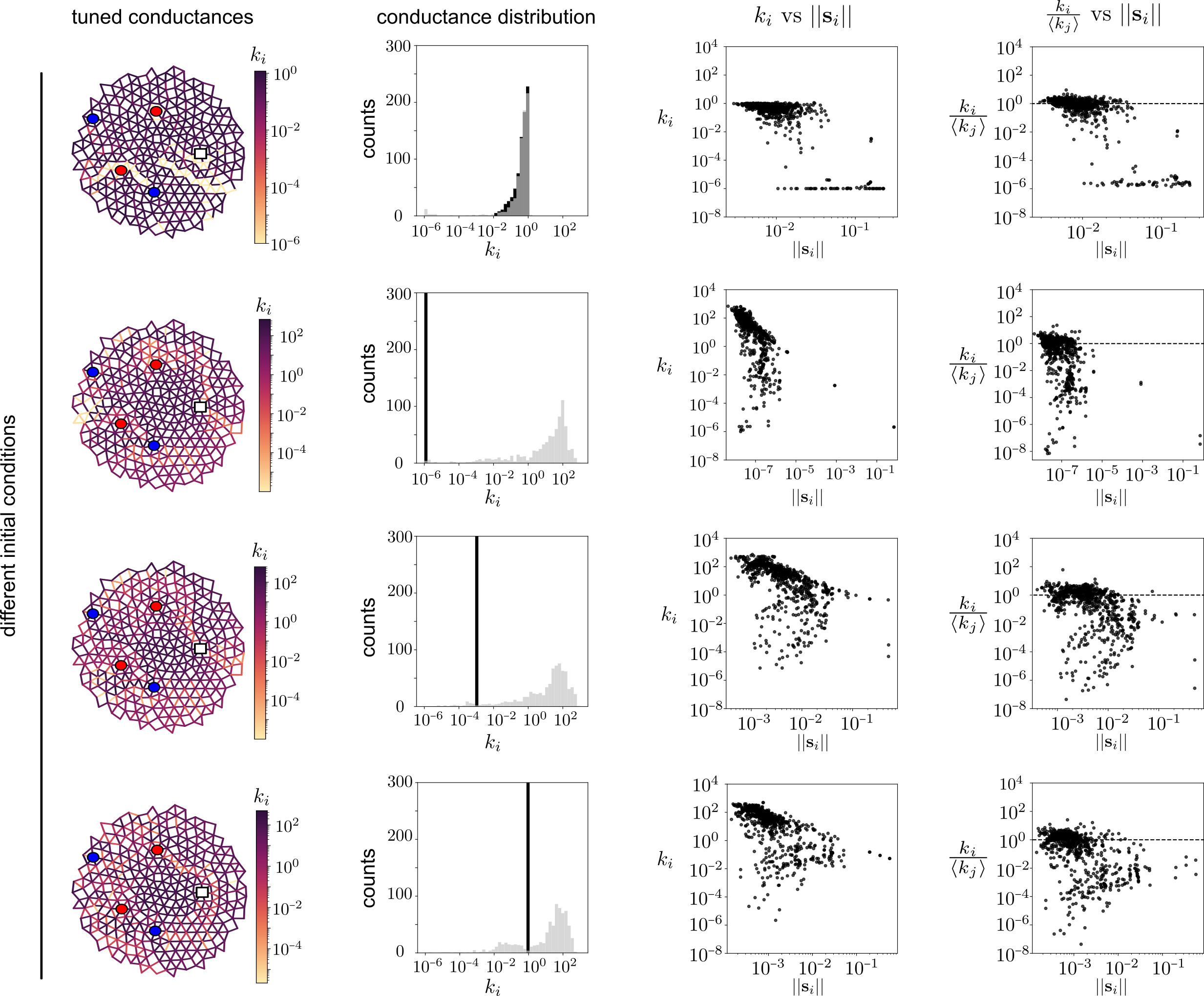}
\caption{Physical properties of  four networks (rows) trained for the same task with different initial conditions.
From left to right: conductance configuration of the trained networks; initial (light gray) and final (black) distribution of conductances; relation between final conductances and susceptibilities; relation between final relative conductances and susceptibilities (horizontal dashed lined set at 1).
}
\label{fig:conductances}
\end{figure*}
One of the hallmarks of physical learning in linear networks is the low dimensionality of the physical response~\cite{stern2021physical}.
Mechanical and resistor networks have been shown to exhibit trained responses that depend only on a handful of physical soft modes, representing a small fraction of all the modes.
This low dimensionality is also suggested by the susceptibility norm in Eq.~\eqref{eq:norm}: the mode contributions decay with increasing magnitude of the eigenvalues.
We explore the dimensionality reduction by determining the number of modes required to define the susceptibility norm.
We define the partial susceptibility as
\begin{equation}
    \mathbf s_i^{(n)} = \sum_{\alpha = 1}^n\frac{1}{\nu_{\alpha}}\left(\mathbf \Delta_i^T\mathbf W_{\alpha}\right) \mathbf W_{\alpha}^T.
\end{equation}
Clearly, we recover the full susceptibility when $n=N+1$, i.e. the total number of modes.
Figure~\ref{fig:size} shows the norm of the partial susceptibility as a function of the number of modes for all four cases studied in the paper.
Aligned with the dimensionality reduction of trained responses, for all cases the norm saturates at values ranging from $n=2$ to $n=10$, showing that most of the physical information is encoded in a few soft modes.

\section{Physical interpretation of role of key edges}
We have established that the susceptibility, a physical quantity devoid of learning information,  correlates positively with the stiff modes of the cost Hessian.
Highly-susceptible edges are therefore important for the functionality of the network.
But how do highly-susceptible edges affect the physical response?
We next study the conductance values of the trained networks and analyze them in relation to their susceptibility values.
\begin{figure*}
\centering
\includegraphics[width=\textwidth]{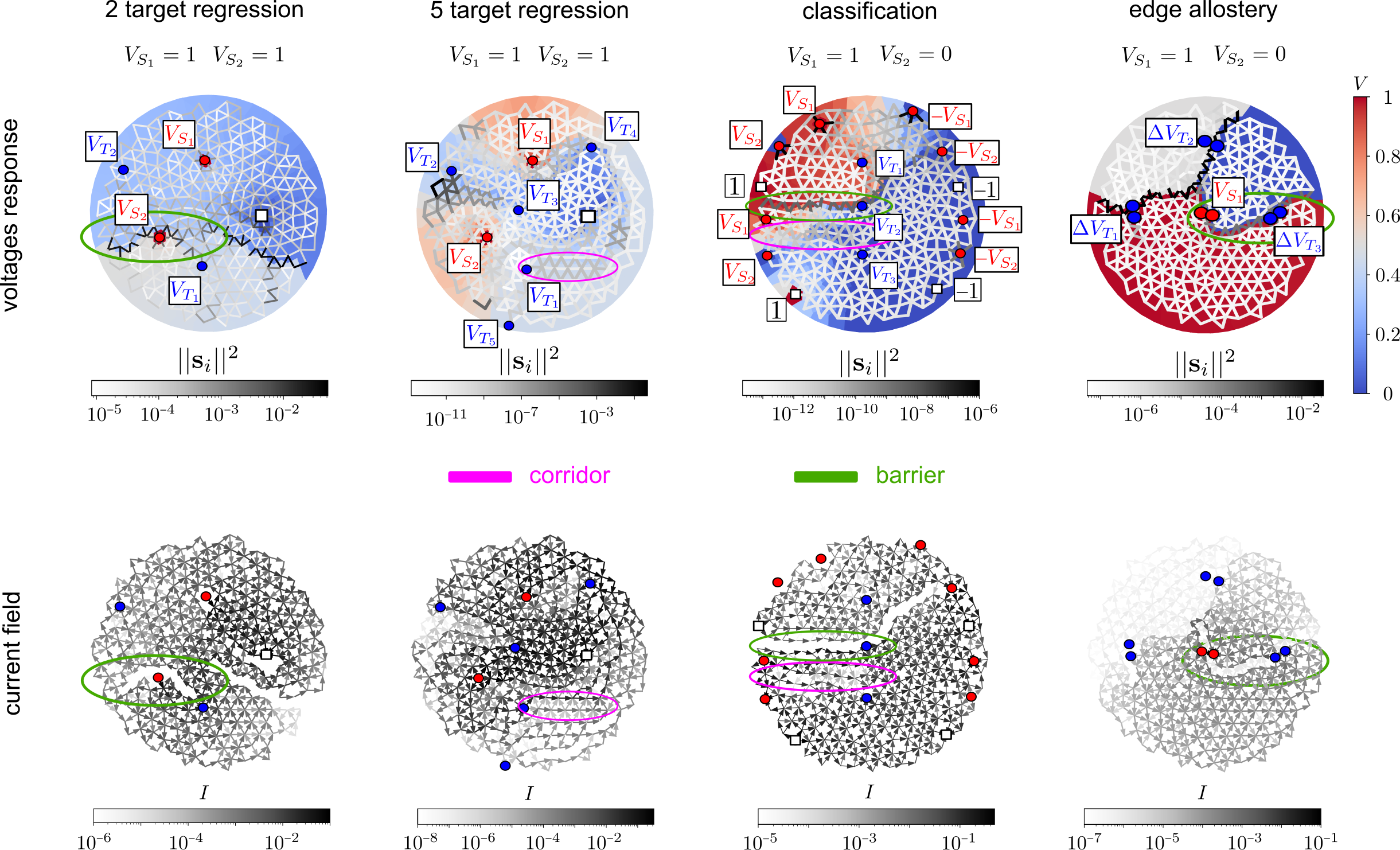}
\caption{Voltage and current responses of the four trained networks considered in the main text (columns).
Top: Voltage response (color) to the  input values shown by the labels.
The edges are colored according to their susceptibility.
Bottom: the associated current response represented as a vector field with orientation indicated by the arrows, and magnitude by the gray scale.
Across all examples, highly-susceptible edges have low relative conductance and therefore block the current.
We highlight two distinct functional mechanisms of the susceptible edges. We use green to highlight susceptible edges that block the current to concentrate voltage, and use pink to highlight those that create a corridor to direct the current along it.
}
\label{fig:current}
\end{figure*}

We consider the linear regression task of Fig.~\ref{fig:example} with four sets of random initial conditions: uniformly-distributed between $10^{-2}$ and $1$ and sharper distributions peaked at $10^{-6}$, $10^{-3}$, and $10^{-1}$.
The results are shown in fig.~\ref{fig:conductances}:
Different initial configurations lead to qualitatively different final states, each of them performing exactly the same task. The existence of multiple good solutions is a signature of the overparameterization of the networks (first column and second column).
With the exception of the first example (first row), conductance is generically not correlated with the susceptibility (third column).
This observations also holds for all the examples analyzed presented in the main text (see Appendix \ref{sec:susceptibility_conductances}).
The fourth column shows the relation between the \emph{relative conductance} $k_i/\left<k_i\right>$ of each edge with its susceptibility $||\mathbf s_i||^2$, where $\left<k_i\right>$ is the average conductance of the adjacent edges.
While their relation depends on the task and initial condition, one striking feature is common across all the examples: high-susceptibility edges have low relative conductances $k_i/\left<k_i\right><1$, (see Appendix \ref{sec:susceptibility_conductances}) for more examples).
We highlight that this does not need to be the case, as one can imagine a functional network relying on connected segments of very high conductance (pipelines) to accomplish the same task.
The training procedure we use (Coupled Learning) does not appear to generically yield such solutions; how different training protocols might affect the nature of trained solutions is an interesting subject for further study.

In Fig.~\ref{fig:current} we illustrate how edges with high susceptibility and low relative conductance effectively act as current blockers, or barriers, for generic input voltages.
We analyze the four tasks of the main text: linear regression with two targets, with five targets, classification, and trained voltage drops at output edges (different columns).
The first row illustrates the voltage response (background color) for given input voltages and highly susceptible edges (gray scale).
The current field, shown in the second row, shows how current flows from inputs to outputs in the network.
Highly-susceptible edges block the current along them, redirecting it by at least two identifiable mechanisms: acting as a wall to concentrate voltage on either side (highlighted in green), or forming corridors through which the current is concentrated and routed to specific locations (highlighted in pink).

The system places walls and corridors to accomplish the trained task.
Take, for example, the linear regression task for two targets (first row).
A wall separates nodes $T_2$ from $S_2$ but not $S_1$. This is a sensible strategy because $T_2$ is nearly equally distant from both $S_1$ and $S_2$ but was trained for a dependence on $S_2$ that is only half as strong as the dependence on $S_1$ (see eq.~\eqref{eq:task}). 
Similar conclusions can be drawn for the other tasks (Appendix~\ref{sec:susceptibility_conductances}).
These results show that the joint information of susceptibility and conductance can provide physical insight into how the systems accomplish their tasks.

\section{Discussion}
Trained  resistor networks have their learned solutions imprinted into their physical landscapes.
We have shown that the local geometry of the cost landscape around the learned solution, characterized by the cost Hessian, stems from two separable quantities.
The first is the training tensor, $\mathbf L$, which depends solely on the task the system has been trained to perform: the location of input and output nodes, and the input currents corresponding to the training samples.
The second quantity is the susceptibility tensor $\mathbf S$, which depends solely on physical  properties of the network, namely its architecture and the edge conductances.

The explicit decoupling of training (task-dependent) and physical quantities has several consequences.
First, it extends and formalizes the correspondence between cost and physical Hessians~\cite{stern2024hessians}, allowing us to extract the key, highly-susceptible edges.
The key edges are those indicated by the stiff modes of the cost Hessian, and are the most important edges for performing the task. Here we have shown that they tend to have high susceptibilities, so that they can be identified from physical properties of the network alone, without needing to know the task. 

Second, it establishes a fundamental bound on the physical information contained in the cost Hessian.
For linear resistor networks, {\it all } the physical information is encoded in the susceptibility tensor.
For nonlinear networks, all of the physical information up to quadratic order in the physical landscape is encoded in the susceptibility tensor.

We have shown that the susceptibility tensor $\mathbf S$, which can be decomposed into susceptibility vectors $\mathbf s_i$, explains previous observations on physical learning such as the low-dimensional response~\cite{stern2024physical}, the mode correspondence in simple tasks~\cite{stern2024hessians}, and the topological structure of allosteric networks~\cite{rocks2020StructureFunction}.

Third, it provides an experimentally measurable quantity independent of the training details.
Measurements such as the response to localized input currents can be used to extract  $\mathbf s_i$, revealing critical information about how the task is performed.

In short, the double optimization at the heart of physical learning in many systems makes it possible to gain physical insight into how trained tasks are accomplished. Such insight does not exist for artificial neural networks, which do not possess a physical landscape.

While all the derivations were explicitly done in the context of resistor networks, the same results naturally extend to elastic networks, where the response minimizes energy, and in general, to any physical network that minimizes a scalar quantity (a Lyapunov function) and whose response can be approximated by a physical Hessian around a known minimum. 

As an example, elastic networks have been trained to have responses similar to protein allostery, in which strain applied by binding one molecule affects the ability of a second type of molecule to bind to the protein elsewhere. Spring networks have been trained to develop allosteric responses using global gradient descent~\cite{rocks2017designing}, Monte Carlo methods~\cite{yan2017architecture,rouviere2023emergence}, or local learning rules such as directed aging~\cite{hexner2019periodic} and Coupled Learning~\cite{stern2021supervised}. By contrast, \emph{real} proteins have been trained to perform allostery via biological evolution. The generality of our theoretical analysis implies that it may be applied independent of the process by which a system developed the ability to perform a task. 
This generality suggests that the connection~\cite{husain2020physical} between slow physical modes of proteins and underlying nonlinearities that correlate non-pairwise additive effects of mutations of amino acids in the sequence (global epistasis) can be generalized to gain physical insight into global epistasis even in cases where allostery is not characterized by a single slow mode. 

Furthermore, our methods should yield potentially useful insight into \emph{how} proteins perform allosteric tasks. Using persistent homology, a topological data analysis tool, it has been shown that the functional response of trained allosteric networks can often be described by robust macroscopic sectors highlighting large-scale structures in the network~\cite{rocks2024topological} akin to functional regions observed in allosteric proteins~\cite{halabi2024protein}.
The persistent homology analysis depends on both training (input nodes and values) and physical (response) information.
We have shown that similar topological structures can be obtained with physical information alone, through the susceptibility tensor, in a closely related system. An important direction for future work is to apply our analysis to mechanical networks with allostery to extract key edges and compare to persistent homology results.

More generally, our results pave the way towards understanding and interpreting how collective trained behavior emerges. Here we have studied linear networks, for which the physical landscape is completely convex and possesses only one minimum.
Biological, physical, and artificial systems, however, heavily rely on non-linearities to achieve more and more complex tasks~\cite{alim2017mechanism,rouviere2023emergence,hexner2019effect,anisetti2022frequency,hornik1989multilayer}.
This is the case, for example, of elastic networks with multistable allostery~\cite{rouviere2023emergence} and also resistor networks that learn non-linear regression and classification~\cite{dillavou2024machine,stern2024training}.
For such systems, the physical landscape is decorated with several minima, with each minimum characterized by different susceptibilities. Understanding how distributions and correlations among edge susceptibilities might vary among minima is an important area of future work, akin to the study of statistical comparisons of the properties of different jammed minima in sphere packings~\cite{liu2010jamming}.

\section{Acknowledgements}
This work was supported by DOE Basic Energy Sciences through grant DE-SC0020963 (MG,FM,MS), the UPenn NSF NRT DGE-2152205 (FM) and the Simons Foundation through Investigator grant \#327939 to AJL. MS was also supported by NIH CRCNS grant 1R01MH125544-01 and NSF grant CISE 2212519.

\appendix
\section{$\mathbf L$-$\mathbf S$ decomposition for general cost functions}
\label{sec:A1}
The cost function $C$ can be written as
\begin{equation}
C=\frac{1}{2}\sum_{r=1}^{R} \bm c_r^T\bm c_r,
\label{eq:cost_general}
\end{equation}
where $\bm c_r=\bm c_r(Q_r\mathbf V_r^F)$ are generic (non-linear) differentiable constraints depending on the voltages at the output nodes, $Q_r\mathbf V_r^F$.
For fully trained networks, $C=0$, and the cost hessian is given by
\begin{equation}
    \mathcal H_{ij} = \frac{\partial ^2 C}{\partial k_i\partial k_j}= \sum_{r=1}^R \frac{\partial \bm c_{r}^T}{\partial k_i}\frac{\partial \bm c_{r}}{\partial k_j},
\end{equation}
where,
\begin{equation}
    \frac{\partial \bm c_r}{\partial k_i}= {{\bm c}'_r}^T Q_r\frac{\partial \mathbf V^F}{\partial k_i}= 
    {{\bm c}'_r}^T Q_r\frac{\partial \mathbf H^{-1}}{\partial k_i} \mathbf I_r,
\end{equation}
and $\mathbf c'_r =\mathbf c'_r(Q_r\mathbf V_r^F)$ is the derivative of the constraint with respect to its arguments evaluated at the output voltage nodes.
In the case of the MSE cost used in the main text, this last term is equal to a vector of ones.

Using the last two equations, we can write the cost hessian as
\begin{equation}
\mathcal{H}_{ij} = \sum_{r=1}^R   \mathbf I_r^T\frac{\partial \mathbf H^{-1}}{\partial k_j} Q_r^T\bm c'_r{\bm c'}_r^TQ_r \frac{\partial \mathbf H^{-1}}{\partial k_i} \mathbf I_r,
\end{equation}
which can be again be split into a learning $\mathbf L$ and susceptibility tensor $\mathbf S$:
\begin{equation}
    \mathcal H_{ij}= \mathbf L\fourvdots\mathbf S_{ij}=\sum_{a,b,c,d}\mathbf L_{abcd}\mathbf S_{ij,abcd},
\end{equation}
with
\begin{align}
\mathbf L &= \sum_{r=1}^R\mathbf I_r^T\otimes Q_r^T\bm c'_r{\bm c'}_r^TQ_r\otimes \mathbf I_r,\\
\mathbf S_{ij} &= \frac{\partial\mathbf H ^{-1}}{\partial k_i} \otimes \frac{\partial\mathbf H ^{-1}}{\partial k_j}.
\end{align}

As expected, the susceptibility tensor is not modified by the specifics of the cost function.
\begin{figure*}
\centering
\includegraphics[width=\textwidth]{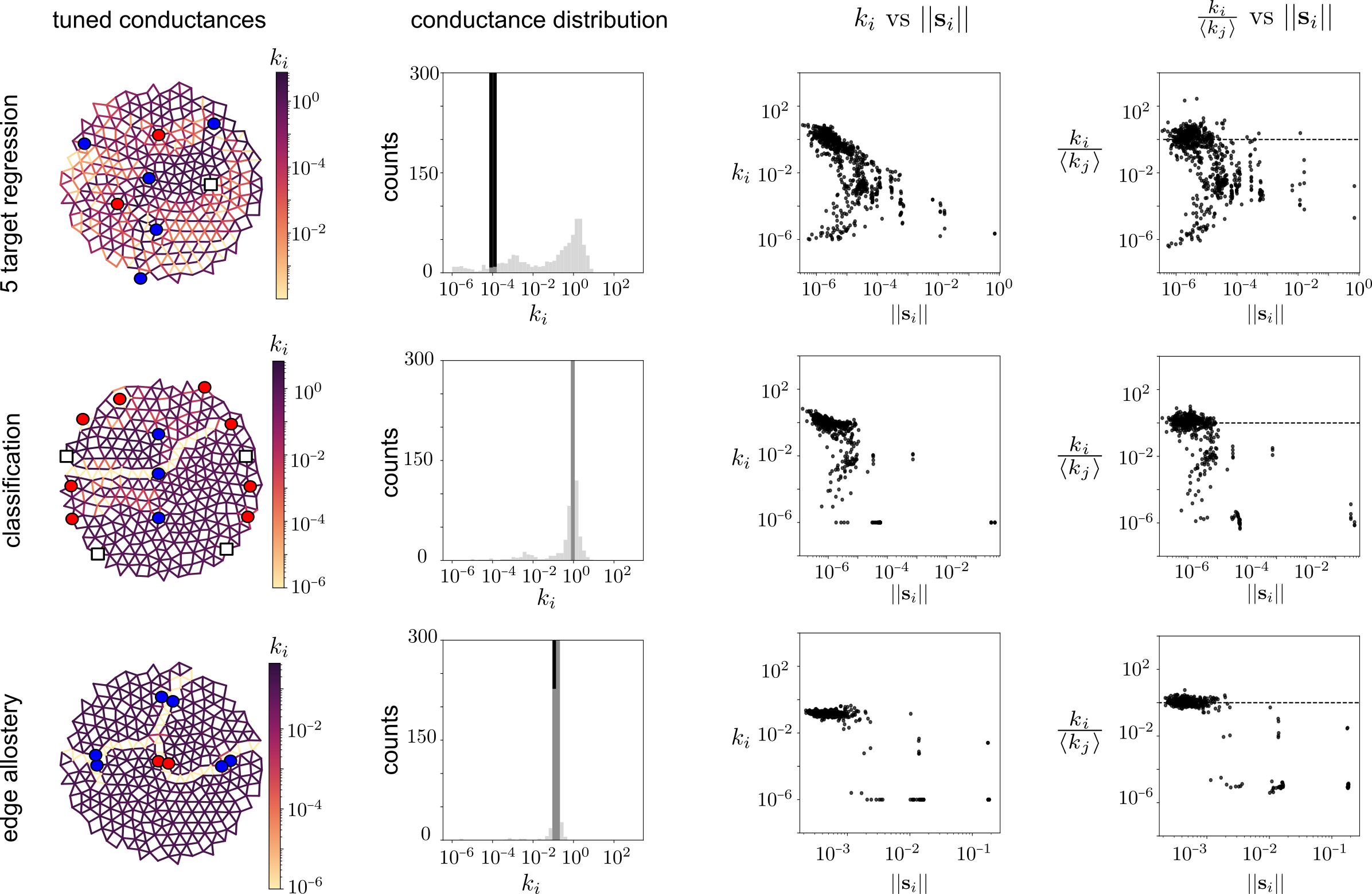}
\caption{Physical properties of three networks (rows) trained for the different tasks. From top to bottom we have circuits trained for: linear regression response at five targets, classification of three clusters (both shown in shown in Fig. \ref{fig:stiff_suscept_regression2_class}), and edge allostery, shown in Fig. \ref{fig:edge_allo}.
From left to right: conductance configuration of the trained networks; initial (light gray) and final (black) distribution of conductances; relation between final conductances and susceptibilities; relation between final relative conductances and susceptibilities (horizontal dashed lined set at 1). 
}
\label{fig:conductances_Appendix}
\end{figure*}
\section{Comparison between persistent sectors and susceptibility in allosteric networks}
\label{sec:PH}
\begin{figure*}
\centering
\includegraphics[width=0.65\textwidth]{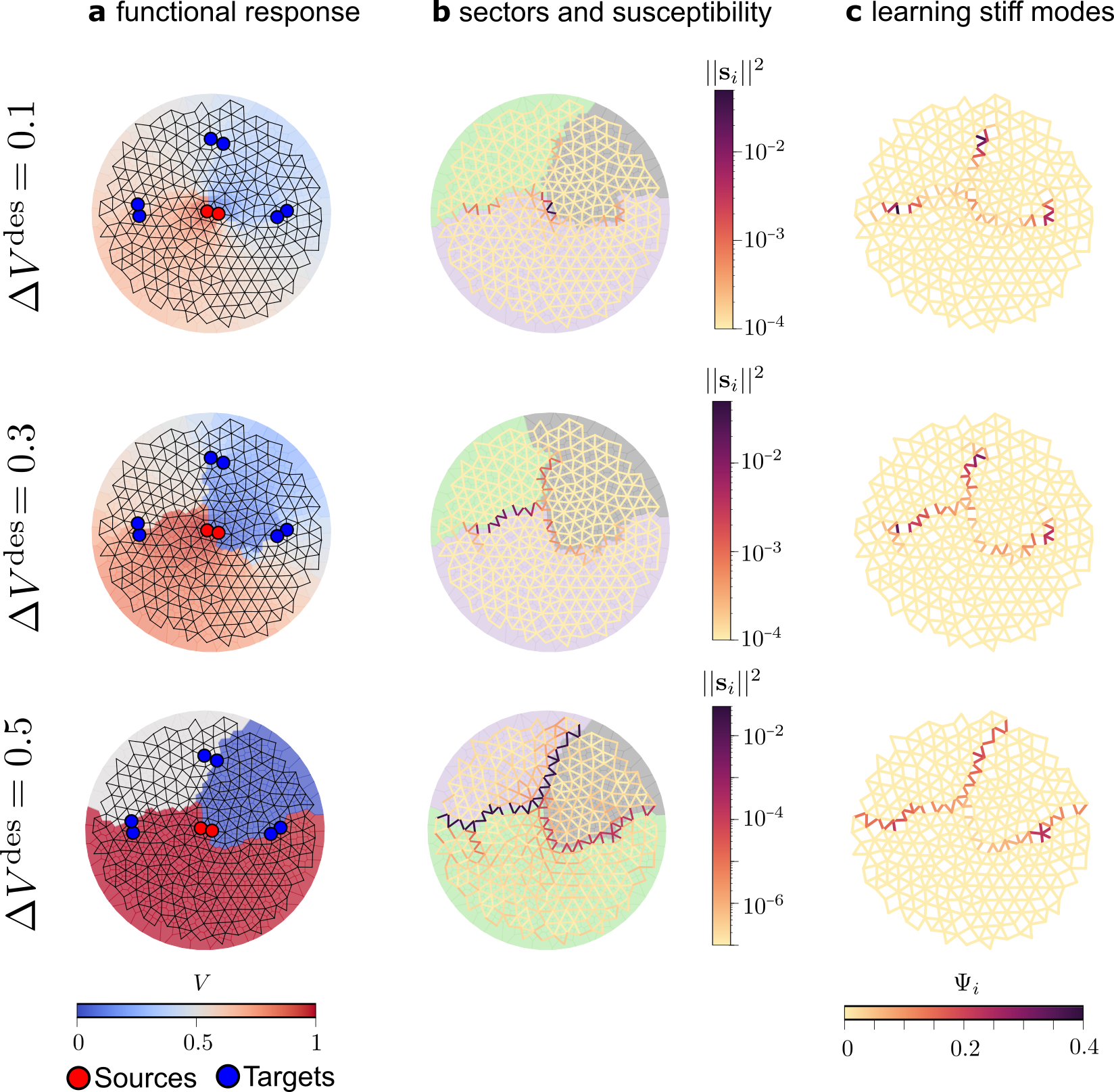}
\caption{
From top to bottom, each row corresponds to the same circuit trained for an allosteric response of increasing strength $\Delta V^{\text{des}}$ at three target sites (blue) in response to source voltages 0 and 1 (red). 
{\bf a.} Voltage response to inputs 0 and 1 (red).
Stronger responses showcase a clearer partitioning of the voltage field. 
{\bf b.} Persistent homology provides a topological partition of the network into distinct sectors (background color), based on the voltage response.
Edges are colored according to their susceptibility, delimiting the topological sectors as $\Delta V^{\text{des}}$ increases. 
{\bf c.} The three stiff modes of the cost hessian closely match the highly susceptible edges, in spite of the topological sectors.
Edges are colored by the sum of stiff modes ($\sum_{\rho}|\Psi_{\rho,i}|$) after normalization.}
\label{fig:PH_appendix}
\end{figure*}

Here we provide further evidence that the sectors obtained by persistence homology in allosteric networks are readily captured by the pattern of susceptibilities, and that they correspond to the stiff modes of the cost hessian.
We consider the same task as in the main text: two inputs nodes with voltages 0 and 1, and three target edges trained for the same voltage drop value $\Delta V^{\text{des}}$.
In addition to the original task shown in fig.~\ref{fig:edge_allo} ($\Delta V^{\text{des}}=0.5$), we consider two more subtle cases: $\Delta V^{\text{des}}=0.3$ and $\Delta V^{\text{des}}=0.1$, which require much less training but leave weaker imprints into the cost hessian.
The voltage response field shows sharper variations as $\Delta V^{\text{des}}$ increases, fig.~\ref{fig:PH_appendix}a, leading to clear partitions, or sectors, captured by persistent homology~\cite{rocks2021hidden}, fig.~\ref{fig:PH_appendix}b.
As $\Delta V^{\text{des}}$ decreases, the susceptible edges have a smaller overlap with the boundaries of the persistent sectors, yet they accurately capture the stiff modes of the cost hessian, fig.~\ref{fig:PH_appendix}c.
For details of the persistent homology algorithm see \cite{rocks2021hidden}.

\section{Susceptibilities and conductances of remaining examples}
\label{sec:susceptibility_conductances}
In figure~\ref{fig:conductances_Appendix} we show the distribution of conductances and edge susceptibilities for the circuits in figures \ref{fig:stiff_suscept_regression2_class}a, \ref{fig:stiff_suscept_regression2_class}e, and  \ref{fig:edge_allo}.
For all these cases, the same general features mentioned in the main text hold: high-susceptible edges generically have low relative conductances, while the conductance itself is not indicative of the susceptibility values.

\section{Training protocol}
\label{sec:A2}
All the networks were trained using Coupled Learning~\cite{stern2021physical}, a contrastive local learning rule with two hyperparameters: the learning rate $\alpha$ and the nudge parameter $\eta$ (for details of the training scheme, see~\cite{stern2021physical}).

The first linear regression circuit (fig.~\ref{fig:example}) was trained for $\approx 10^6$ 
iterations, each iteration batching 30 training samples, using $\alpha = 10^{-4}$, $\eta = 10^{-2}$, and initial conductances drawn from a uniform distribution in  $[0.01,1]$, reaching a final MSE cost $C\approx10^{-5}$. 

The input data consisted of 112 pairs of voltages $(V_{S_1}, V_{S_2})$ with corresponding outputs given by the linear relation of eq.~\eqref{eq:task}. 

The second linear regression circuit (fig. \ref{fig:stiff_suscept_regression2_class}a) was trained for $\approx 3\times 10^5$
iterations,  each iteration batching 30 training samples, 
using $\alpha = 10^{-2}$, $\eta = 10^{-3}$, and initial conductances drawn from a uniform distribution in  $[0.0001,0.00011]$, reaching a final MSE cost $C\approx10^{-8}$. 
The input data consisted of 112 pairs of voltages $(V_{S_1}, V_{S_2})$ with corresponding outputs given by the linear relation of eq.~\eqref{eq:5T_regression_coeffs}.

The classification circuit (fig.~\ref{fig:stiff_suscept_regression2_class}e) was trained for $\approx 2\times 10^4$ 
iterations, each iteration batching 30 training samples, 
using $\alpha = 10^{-1}$, $\eta = 10^{-2}$, and initial conductances all equal to $1$, reaching a final cosine similarity cost $C\approx0.05$, train accuracy of $99\%$ (120 points), and test accuracy of $100\%$ (30 points).
The input data consisted of three clusters of points, one per class, generated from the following normal distributions: 
\begin{align}
&\mathcal{N}\bigg([2,2],\begin{pmatrix}1&0.5\\0.5&1\end{pmatrix}\bigg),\\
&\mathcal{N}\bigg([7,7],\begin{pmatrix}1&-0.5\\-0.5&1\end{pmatrix}\bigg),\\
&\mathcal{N}\bigg([1,8],\begin{pmatrix}1&0\\0&1\end{pmatrix}\bigg).
\end{align}

The allosteric circuit (fig.~\ref{fig:edge_allo}) was trained for $ 5\times 10^4$ iterations, no batching,
using $\alpha = 10^{-2}$, $\eta = 10^{-3}$, and initial conductances were drawn from a uniform distribution in $[0.1,0.2]$, reaching a final MSE cost $C\approx 1\times 10^{-5}$.
The input data is $V_{S_1}=1, V_{S_2}=0$, with corresponding output $\Delta V^{\text{des}}=0.5$.  
The additional two allosteric circuits in fig.~\ref{fig:PH_appendix} were trained for $5\times10^4$ iterations, no batching, using $\alpha = 10^{-2}$, $\eta = 10^{-3}$ and initial conductances were drawn from a uniform distribution in $[0.1,0.2]$, reaching final MSE costs of $C\approx2\times 10^{-8}$ ($\Delta V^{\text{des}}=0.1$) and $C\approx 7\times 10^{-10}$ ($\Delta V^{\text{des}}=0.3$).

\section{Susceptibility for weakly trained networks}
\label{sec:task_strength}
\begin{figure*}
\centering
\includegraphics[width=0.82\textwidth]{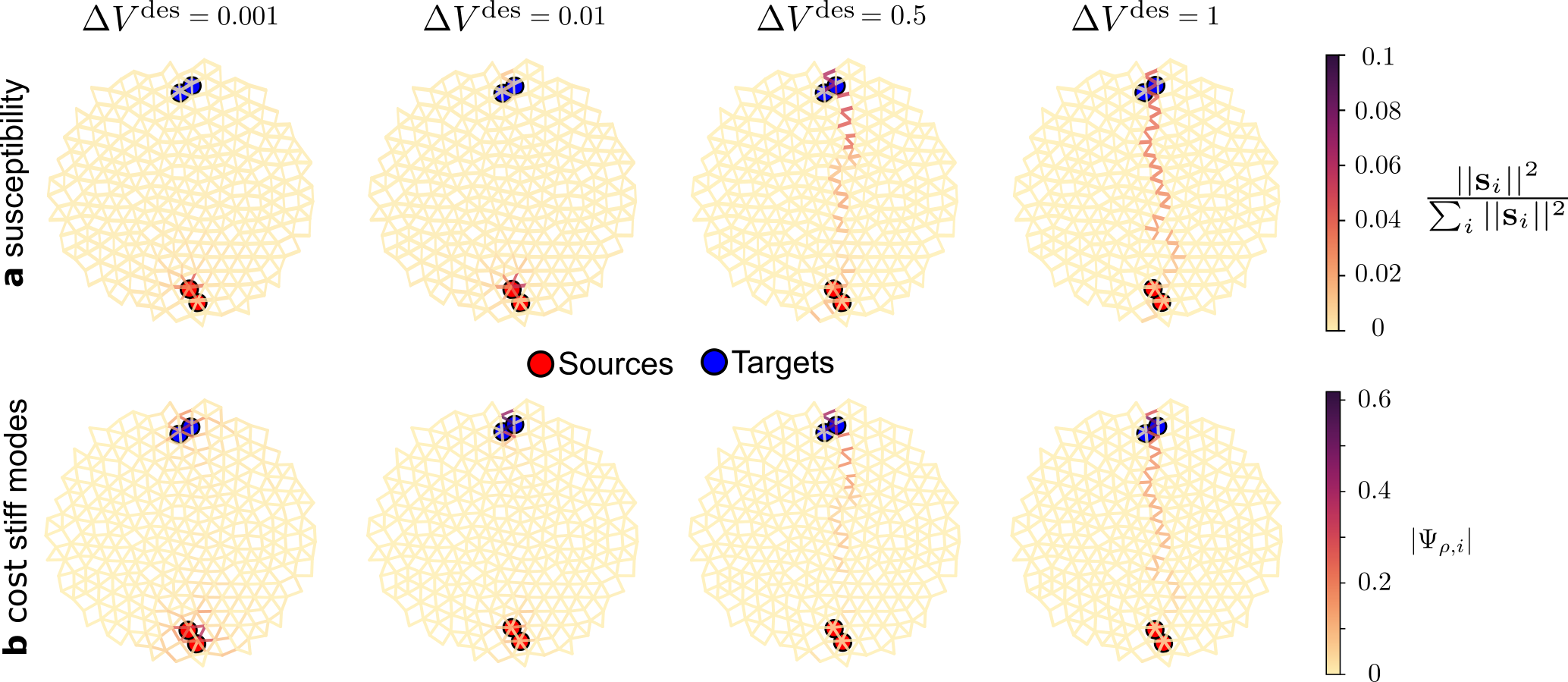}
\caption{From left to right, the normalized susceptibility (top) for increasing allosteric coupling $\Delta V^{\text{des}}$.
Most of the highly susceptible edges correspond to the ones captured by the single stiff mode of the cost hessian (bottom).}
\label{fig:task_strength_appendix}
\end{figure*}

Figure~\ref{fig:task_strength_appendix} shows how the susceptibility depends on the task strength.
We consider an allosteric task with two input voltages (0 and 1), and one edge target with increasing desired voltage drop $\Delta V^{\text{des}}= 0.001,\; 0.01,\; 0.5,\; 1$.
The lower the desired voltage drop, the easier the task and less training steps are required.
In fig.~\ref{fig:task_strength_appendix} we show that despite the weak signals, the susceptibilities still capture the relevant edges highlighted by the stiff mode of the cost hessian.

\bibliography{apssamp}

\begin{thebibliography}{35}%
\makeatletter
\providecommand \@ifxundefined [1]{%
 \@ifx{#1\undefined}
}%
\providecommand \@ifnum [1]{%
 \ifnum #1\expandafter \@firstoftwo
 \else \expandafter \@secondoftwo
 \fi
}%
\providecommand \@ifx [1]{%
 \ifx #1\expandafter \@firstoftwo
 \else \expandafter \@secondoftwo
 \fi
}%
\providecommand \natexlab [1]{#1}%
\providecommand \enquote  [1]{``#1''}%
\providecommand \bibnamefont  [1]{#1}%
\providecommand \bibfnamefont [1]{#1}%
\providecommand \citenamefont [1]{#1}%
\providecommand \href@noop [0]{\@secondoftwo}%
\providecommand \href [0]{\begingroup \@sanitize@url \@href}%
\providecommand \@href[1]{\@@startlink{#1}\@@href}%
\providecommand \@@href[1]{\endgroup#1\@@endlink}%
\providecommand \@sanitize@url [0]{\catcode `\\12\catcode `\$12\catcode
  `\&12\catcode `\#12\catcode `\^12\catcode `\_12\catcode `\%12\relax}%
\providecommand \@@startlink[1]{}%
\providecommand \@@endlink[0]{}%
\providecommand \url  [0]{\begingroup\@sanitize@url \@url }%
\providecommand \@url [1]{\endgroup\@href {#1}{\urlprefix }}%
\providecommand \urlprefix  [0]{URL }%
\providecommand \Eprint [0]{\href }%
\providecommand \doibase [0]{https://doi.org/}%
\providecommand \selectlanguage [0]{\@gobble}%
\providecommand \bibinfo  [0]{\@secondoftwo}%
\providecommand \bibfield  [0]{\@secondoftwo}%
\providecommand \translation [1]{[#1]}%
\providecommand \BibitemOpen [0]{}%
\providecommand \bibitemStop [0]{}%
\providecommand \bibitemNoStop [0]{.\EOS\space}%
\providecommand \EOS [0]{\spacefactor3000\relax}%
\providecommand \BibitemShut  [1]{\csname bibitem#1\endcsname}%
\let\auto@bib@innerbib\@empty
\bibitem [{\citenamefont {Mehta}\ \emph {et~al.}(2019)\citenamefont {Mehta},
  \citenamefont {Bukov}, \citenamefont {Wang}, \citenamefont {Day},
  \citenamefont {Richardson}, \citenamefont {Fisher},\ and\ \citenamefont
  {Schwab}}]{mehta2019high}%
  \BibitemOpen
  \bibfield  {author} {\bibinfo {author} {\bibfnamefont {P.}~\bibnamefont
  {Mehta}}, \bibinfo {author} {\bibfnamefont {M.}~\bibnamefont {Bukov}},
  \bibinfo {author} {\bibfnamefont {C.-H.}\ \bibnamefont {Wang}}, \bibinfo
  {author} {\bibfnamefont {A.~G.}\ \bibnamefont {Day}}, \bibinfo {author}
  {\bibfnamefont {C.}~\bibnamefont {Richardson}}, \bibinfo {author}
  {\bibfnamefont {C.~K.}\ \bibnamefont {Fisher}},\ and\ \bibinfo {author}
  {\bibfnamefont {D.~J.}\ \bibnamefont {Schwab}},\ }\bibfield  {title}
  {\bibinfo {title} {A high-bias, low-variance introduction to machine learning
  for physicists},\ }\href@noop {} {\bibfield  {journal} {\bibinfo  {journal}
  {Physics reports}\ } (\bibinfo {year} {2019})}\BibitemShut {NoStop}%
\bibitem [{\citenamefont {Sagun}\ \emph {et~al.}(2016)\citenamefont {Sagun},
  \citenamefont {Bottou},\ and\ \citenamefont {LeCun}}]{sagun2016eigenvalues}%
  \BibitemOpen
  \bibfield  {author} {\bibinfo {author} {\bibfnamefont {L.}~\bibnamefont
  {Sagun}}, \bibinfo {author} {\bibfnamefont {L.}~\bibnamefont {Bottou}},\ and\
  \bibinfo {author} {\bibfnamefont {Y.}~\bibnamefont {LeCun}},\ }\bibfield
  {title} {\bibinfo {title} {Eigenvalues of the hessian in deep learning:
  Singularity and beyond},\ }\href@noop {} {\bibfield  {journal} {\bibinfo
  {journal} {arXiv preprint arXiv:1611.07476}\ } (\bibinfo {year}
  {2016})}\BibitemShut {NoStop}%
\bibitem [{\citenamefont {Sagun}\ \emph {et~al.}(2017)\citenamefont {Sagun},
  \citenamefont {Evci}, \citenamefont {Guney}, \citenamefont {Dauphin},\ and\
  \citenamefont {Bottou}}]{sagun2017empirical}%
  \BibitemOpen
  \bibfield  {author} {\bibinfo {author} {\bibfnamefont {L.}~\bibnamefont
  {Sagun}}, \bibinfo {author} {\bibfnamefont {U.}~\bibnamefont {Evci}},
  \bibinfo {author} {\bibfnamefont {V.~U.}\ \bibnamefont {Guney}}, \bibinfo
  {author} {\bibfnamefont {Y.}~\bibnamefont {Dauphin}},\ and\ \bibinfo {author}
  {\bibfnamefont {L.}~\bibnamefont {Bottou}},\ }\bibfield  {title} {\bibinfo
  {title} {Empirical analysis of the hessian of over-parametrized neural
  networks},\ }\href@noop {} {\bibfield  {journal} {\bibinfo  {journal} {arXiv
  preprint arXiv:1706.04454}\ } (\bibinfo {year} {2017})}\BibitemShut {NoStop}%
\bibitem [{\citenamefont {Dauphin}\ \emph {et~al.}(2014)\citenamefont
  {Dauphin}, \citenamefont {Pascanu}, \citenamefont {Gulcehre}, \citenamefont
  {Cho}, \citenamefont {Ganguli},\ and\ \citenamefont
  {Bengio}}]{dauphin2014identifying}%
  \BibitemOpen
  \bibfield  {author} {\bibinfo {author} {\bibfnamefont {Y.~N.}\ \bibnamefont
  {Dauphin}}, \bibinfo {author} {\bibfnamefont {R.}~\bibnamefont {Pascanu}},
  \bibinfo {author} {\bibfnamefont {C.}~\bibnamefont {Gulcehre}}, \bibinfo
  {author} {\bibfnamefont {K.}~\bibnamefont {Cho}}, \bibinfo {author}
  {\bibfnamefont {S.}~\bibnamefont {Ganguli}},\ and\ \bibinfo {author}
  {\bibfnamefont {Y.}~\bibnamefont {Bengio}},\ }\bibfield  {title} {\bibinfo
  {title} {Identifying and attacking the saddle point problem in
  high-dimensional non-convex optimization},\ }\href@noop {} {\bibfield
  {journal} {\bibinfo  {journal} {Advances in neural information processing
  systems}\ }\textbf {\bibinfo {volume} {27}} (\bibinfo {year}
  {2014})}\BibitemShut {NoStop}%
\bibitem [{\citenamefont {Sabanayagam}\ \emph {et~al.}(2023)\citenamefont
  {Sabanayagam}, \citenamefont {Behrens}, \citenamefont {Adomaityte},\ and\
  \citenamefont {Dawid}}]{sabanayagam2023unveiling}%
  \BibitemOpen
  \bibfield  {author} {\bibinfo {author} {\bibfnamefont {M.}~\bibnamefont
  {Sabanayagam}}, \bibinfo {author} {\bibfnamefont {F.}~\bibnamefont
  {Behrens}}, \bibinfo {author} {\bibfnamefont {U.}~\bibnamefont
  {Adomaityte}},\ and\ \bibinfo {author} {\bibfnamefont {A.}~\bibnamefont
  {Dawid}},\ }\bibfield  {title} {\bibinfo {title} {Unveiling the hessian's
  connection to the decision boundary},\ }\href@noop {} {\bibfield  {journal}
  {\bibinfo  {journal} {arXiv preprint arXiv:2306.07104}\ } (\bibinfo {year}
  {2023})}\BibitemShut {NoStop}%
\bibitem [{\citenamefont {Rocks}\ \emph {et~al.}(2017)\citenamefont {Rocks},
  \citenamefont {Pashine}, \citenamefont {Bischofberger}, \citenamefont
  {Goodrich}, \citenamefont {Liu},\ and\ \citenamefont
  {Nagel}}]{rocks2017designing}%
  \BibitemOpen
  \bibfield  {author} {\bibinfo {author} {\bibfnamefont {J.~W.}\ \bibnamefont
  {Rocks}}, \bibinfo {author} {\bibfnamefont {N.}~\bibnamefont {Pashine}},
  \bibinfo {author} {\bibfnamefont {I.}~\bibnamefont {Bischofberger}}, \bibinfo
  {author} {\bibfnamefont {C.~P.}\ \bibnamefont {Goodrich}}, \bibinfo {author}
  {\bibfnamefont {A.~J.}\ \bibnamefont {Liu}},\ and\ \bibinfo {author}
  {\bibfnamefont {S.~R.}\ \bibnamefont {Nagel}},\ }\bibfield  {title} {\bibinfo
  {title} {Designing allostery-inspired response in mechanical networks},\
  }\href@noop {} {\bibfield  {journal} {\bibinfo  {journal} {Proceedings of the
  National Academy of Sciences}\ }\textbf {\bibinfo {volume} {114}},\ \bibinfo
  {pages} {2520} (\bibinfo {year} {2017})}\BibitemShut {NoStop}%
\bibitem [{\citenamefont {Pinson}\ \emph {et~al.}(2017)\citenamefont {Pinson},
  \citenamefont {Stern}, \citenamefont {Carruthers~Ferrero}, \citenamefont
  {Witten}, \citenamefont {Chen},\ and\ \citenamefont
  {Murugan}}]{pinson2017self}%
  \BibitemOpen
  \bibfield  {author} {\bibinfo {author} {\bibfnamefont {M.~B.}\ \bibnamefont
  {Pinson}}, \bibinfo {author} {\bibfnamefont {M.}~\bibnamefont {Stern}},
  \bibinfo {author} {\bibfnamefont {A.}~\bibnamefont {Carruthers~Ferrero}},
  \bibinfo {author} {\bibfnamefont {T.~A.}\ \bibnamefont {Witten}}, \bibinfo
  {author} {\bibfnamefont {E.}~\bibnamefont {Chen}},\ and\ \bibinfo {author}
  {\bibfnamefont {A.}~\bibnamefont {Murugan}},\ }\bibfield  {title} {\bibinfo
  {title} {Self-folding origami at any energy scale},\ }\href@noop {}
  {\bibfield  {journal} {\bibinfo  {journal} {Nature communications}\ }\textbf
  {\bibinfo {volume} {8}},\ \bibinfo {pages} {15477} (\bibinfo {year}
  {2017})}\BibitemShut {NoStop}%
\bibitem [{\citenamefont {Hexner}\ \emph
  {et~al.}(2020{\natexlab{a}})\citenamefont {Hexner}, \citenamefont {Pashine},
  \citenamefont {Liu},\ and\ \citenamefont {Nagel}}]{hexner2019effect}%
  \BibitemOpen
  \bibfield  {author} {\bibinfo {author} {\bibfnamefont {D.}~\bibnamefont
  {Hexner}}, \bibinfo {author} {\bibfnamefont {N.}~\bibnamefont {Pashine}},
  \bibinfo {author} {\bibfnamefont {A.~J.}\ \bibnamefont {Liu}},\ and\ \bibinfo
  {author} {\bibfnamefont {S.~R.}\ \bibnamefont {Nagel}},\ }\bibfield  {title}
  {\bibinfo {title} {Effect of directed aging on nonlinear elasticity and
  memory formation in a material},\ }\href@noop {} {\bibfield  {journal}
  {\bibinfo  {journal} {Physical Review Research}\ }\textbf {\bibinfo {volume}
  {2}},\ \bibinfo {pages} {043231} (\bibinfo {year}
  {2020}{\natexlab{a}})}\BibitemShut {NoStop}%
\bibitem [{\citenamefont {Pashine}(2021)}]{pashine2021local}%
  \BibitemOpen
  \bibfield  {author} {\bibinfo {author} {\bibfnamefont {N.}~\bibnamefont
  {Pashine}},\ }\bibfield  {title} {\bibinfo {title} {Local rules for
  fabricating allosteric networks},\ }\href@noop {} {\bibfield  {journal}
  {\bibinfo  {journal} {Physical Review Materials}\ }\textbf {\bibinfo {volume}
  {5}},\ \bibinfo {pages} {065607} (\bibinfo {year} {2021})}\BibitemShut
  {NoStop}%
\bibitem [{\citenamefont {Arinze}\ \emph {et~al.}(2023)\citenamefont {Arinze},
  \citenamefont {Stern}, \citenamefont {Nagel},\ and\ \citenamefont
  {Murugan}}]{arinze2023learning}%
  \BibitemOpen
  \bibfield  {author} {\bibinfo {author} {\bibfnamefont {C.}~\bibnamefont
  {Arinze}}, \bibinfo {author} {\bibfnamefont {M.}~\bibnamefont {Stern}},
  \bibinfo {author} {\bibfnamefont {S.~R.}\ \bibnamefont {Nagel}},\ and\
  \bibinfo {author} {\bibfnamefont {A.}~\bibnamefont {Murugan}},\ }\bibfield
  {title} {\bibinfo {title} {Learning to self-fold at a bifurcation},\ }\href
  {https://doi.org/10.1103/PhysRevE.107.025001} {\bibfield  {journal} {\bibinfo
   {journal} {Phys. Rev. E}\ }\textbf {\bibinfo {volume} {107}},\ \bibinfo
  {pages} {025001} (\bibinfo {year} {2023})}\BibitemShut {NoStop}%
\bibitem [{\citenamefont {Stern}\ \emph {et~al.}(2021)\citenamefont {Stern},
  \citenamefont {Hexner}, \citenamefont {Rocks},\ and\ \citenamefont
  {Liu}}]{stern2021supervised}%
  \BibitemOpen
  \bibfield  {author} {\bibinfo {author} {\bibfnamefont {M.}~\bibnamefont
  {Stern}}, \bibinfo {author} {\bibfnamefont {D.}~\bibnamefont {Hexner}},
  \bibinfo {author} {\bibfnamefont {J.~W.}\ \bibnamefont {Rocks}},\ and\
  \bibinfo {author} {\bibfnamefont {A.~J.}\ \bibnamefont {Liu}},\ }\bibfield
  {title} {\bibinfo {title} {Supervised learning in physical networks: From
  machine learning to learning machines},\ }\href@noop {} {\bibfield  {journal}
  {\bibinfo  {journal} {Physical Review X}\ }\textbf {\bibinfo {volume} {11}},\
  \bibinfo {pages} {021045} (\bibinfo {year} {2021})}\BibitemShut {NoStop}%
\bibitem [{\citenamefont {Dillavou}\ \emph {et~al.}(2022)\citenamefont
  {Dillavou}, \citenamefont {Stern}, \citenamefont {Liu},\ and\ \citenamefont
  {Durian}}]{dillavou2022demonstration}%
  \BibitemOpen
  \bibfield  {author} {\bibinfo {author} {\bibfnamefont {S.}~\bibnamefont
  {Dillavou}}, \bibinfo {author} {\bibfnamefont {M.}~\bibnamefont {Stern}},
  \bibinfo {author} {\bibfnamefont {A.~J.}\ \bibnamefont {Liu}},\ and\ \bibinfo
  {author} {\bibfnamefont {D.~J.}\ \bibnamefont {Durian}},\ }\bibfield  {title}
  {\bibinfo {title} {Demonstration of decentralized physics-driven learning},\
  }\href@noop {} {\bibfield  {journal} {\bibinfo  {journal} {Physical Review
  Applied}\ }\textbf {\bibinfo {volume} {18}},\ \bibinfo {pages} {014040}
  (\bibinfo {year} {2022})}\BibitemShut {NoStop}%
\bibitem [{\citenamefont {Dillavou}\ \emph {et~al.}(2024)\citenamefont
  {Dillavou}, \citenamefont {Beyer}, \citenamefont {Stern}, \citenamefont
  {Liu}, \citenamefont {Miskin},\ and\ \citenamefont
  {Durian}}]{dillavou2024machine}%
  \BibitemOpen
  \bibfield  {author} {\bibinfo {author} {\bibfnamefont {S.}~\bibnamefont
  {Dillavou}}, \bibinfo {author} {\bibfnamefont {B.~D.}\ \bibnamefont {Beyer}},
  \bibinfo {author} {\bibfnamefont {M.}~\bibnamefont {Stern}}, \bibinfo
  {author} {\bibfnamefont {A.~J.}\ \bibnamefont {Liu}}, \bibinfo {author}
  {\bibfnamefont {M.~Z.}\ \bibnamefont {Miskin}},\ and\ \bibinfo {author}
  {\bibfnamefont {D.~J.}\ \bibnamefont {Durian}},\ }\bibfield  {title}
  {\bibinfo {title} {Machine learning without a processor: Emergent learning in
  a nonlinear analog network},\ }\href@noop {} {\bibfield  {journal} {\bibinfo
  {journal} {Proceedings of the National Academy of Sciences}\ }\textbf
  {\bibinfo {volume} {121}},\ \bibinfo {pages} {e2319718121} (\bibinfo {year}
  {2024})}\BibitemShut {NoStop}%
\bibitem [{\citenamefont {Stern}\ and\ \citenamefont
  {Murugan}(2023)}]{stern2022learning}%
  \BibitemOpen
  \bibfield  {author} {\bibinfo {author} {\bibfnamefont {M.}~\bibnamefont
  {Stern}}\ and\ \bibinfo {author} {\bibfnamefont {A.}~\bibnamefont
  {Murugan}},\ }\bibfield  {title} {\bibinfo {title} {Learning without neurons
  in physical systems},\ }\href
  {https://doi.org/10.1146/annurev-conmatphys-040821-113439} {\bibfield
  {journal} {\bibinfo  {journal} {Annual Review of Condensed Matter Physics}\
  }\textbf {\bibinfo {volume} {14}},\ \bibinfo {pages} {417} (\bibinfo {year}
  {2023})}\BibitemShut {NoStop}%
\bibitem [{\citenamefont {Stern}\ \emph
  {et~al.}(2024{\natexlab{a}})\citenamefont {Stern}, \citenamefont {Guzman},
  \citenamefont {Martins}, \citenamefont {Liu},\ and\ \citenamefont
  {Balasubramanian}}]{stern2024hessians}%
  \BibitemOpen
  \bibfield  {author} {\bibinfo {author} {\bibfnamefont {M.}~\bibnamefont
  {Stern}}, \bibinfo {author} {\bibfnamefont {M.}~\bibnamefont {Guzman}},
  \bibinfo {author} {\bibfnamefont {F.}~\bibnamefont {Martins}}, \bibinfo
  {author} {\bibfnamefont {A.~J.}\ \bibnamefont {Liu}},\ and\ \bibinfo {author}
  {\bibfnamefont {V.}~\bibnamefont {Balasubramanian}},\ }\bibfield  {title}
  {\bibinfo {title} {Physical networks become what they learn},\ }\href@noop {}
  {\bibfield  {journal} {\bibinfo  {journal} {arXiv preprint arXiv:2406.09689}\
  } (\bibinfo {year} {2024}{\natexlab{a}})}\BibitemShut {NoStop}%
\bibitem [{\citenamefont {Vadlamani}\ \emph {et~al.}(2020)\citenamefont
  {Vadlamani}, \citenamefont {Xiao},\ and\ \citenamefont
  {Yablonovitch}}]{vadlamani2020physics}%
  \BibitemOpen
  \bibfield  {author} {\bibinfo {author} {\bibfnamefont {S.~K.}\ \bibnamefont
  {Vadlamani}}, \bibinfo {author} {\bibfnamefont {T.~P.}\ \bibnamefont
  {Xiao}},\ and\ \bibinfo {author} {\bibfnamefont {E.}~\bibnamefont
  {Yablonovitch}},\ }\bibfield  {title} {\bibinfo {title} {Physics successfully
  implements lagrange multiplier optimization},\ }\href@noop {} {\bibfield
  {journal} {\bibinfo  {journal} {Proceedings of the National Academy of
  Sciences}\ }\textbf {\bibinfo {volume} {117}},\ \bibinfo {pages} {26639}
  (\bibinfo {year} {2020})}\BibitemShut {NoStop}%
\bibitem [{\citenamefont {Stern}\ \emph
  {et~al.}(2024{\natexlab{b}})\citenamefont {Stern}, \citenamefont {Liu},\ and\
  \citenamefont {Balasubramanian}}]{stern2024physical}%
  \BibitemOpen
  \bibfield  {author} {\bibinfo {author} {\bibfnamefont {M.}~\bibnamefont
  {Stern}}, \bibinfo {author} {\bibfnamefont {A.~J.}\ \bibnamefont {Liu}},\
  and\ \bibinfo {author} {\bibfnamefont {V.}~\bibnamefont {Balasubramanian}},\
  }\bibfield  {title} {\bibinfo {title} {Physical effects of learning},\
  }\href@noop {} {\bibfield  {journal} {\bibinfo  {journal} {Physical Review
  E}\ }\textbf {\bibinfo {volume} {109}},\ \bibinfo {pages} {024311} (\bibinfo
  {year} {2024}{\natexlab{b}})}\BibitemShut {NoStop}%
\bibitem [{\citenamefont {Tlusty}\ \emph {et~al.}(2017)\citenamefont {Tlusty},
  \citenamefont {Libchaber},\ and\ \citenamefont
  {Eckmann}}]{tlusty2017physical}%
  \BibitemOpen
  \bibfield  {author} {\bibinfo {author} {\bibfnamefont {T.}~\bibnamefont
  {Tlusty}}, \bibinfo {author} {\bibfnamefont {A.}~\bibnamefont {Libchaber}},\
  and\ \bibinfo {author} {\bibfnamefont {J.-P.}\ \bibnamefont {Eckmann}},\
  }\bibfield  {title} {\bibinfo {title} {Physical model of the
  genotype-to-phenotype map of proteins},\ }\href@noop {} {\bibfield  {journal}
  {\bibinfo  {journal} {Physical Review X}\ }\textbf {\bibinfo {volume} {7}},\
  \bibinfo {pages} {021037} (\bibinfo {year} {2017})}\BibitemShut {NoStop}%
\bibitem [{\citenamefont {Yan}\ \emph {et~al.}(2017)\citenamefont {Yan},
  \citenamefont {Ravasio}, \citenamefont {Brito},\ and\ \citenamefont
  {Wyart}}]{yan2017architecture}%
  \BibitemOpen
  \bibfield  {author} {\bibinfo {author} {\bibfnamefont {L.}~\bibnamefont
  {Yan}}, \bibinfo {author} {\bibfnamefont {R.}~\bibnamefont {Ravasio}},
  \bibinfo {author} {\bibfnamefont {C.}~\bibnamefont {Brito}},\ and\ \bibinfo
  {author} {\bibfnamefont {M.}~\bibnamefont {Wyart}},\ }\bibfield  {title}
  {\bibinfo {title} {Architecture and coevolution of allosteric materials},\
  }\href@noop {} {\bibfield  {journal} {\bibinfo  {journal} {Proceedings of the
  National Academy of Sciences}\ }\textbf {\bibinfo {volume} {114}},\ \bibinfo
  {pages} {2526} (\bibinfo {year} {2017})}\BibitemShut {NoStop}%
\bibitem [{\citenamefont {Yan}\ \emph {et~al.}(2018)\citenamefont {Yan},
  \citenamefont {Ravasio}, \citenamefont {Brito},\ and\ \citenamefont
  {Wyart}}]{yan2018principles}%
  \BibitemOpen
  \bibfield  {author} {\bibinfo {author} {\bibfnamefont {L.}~\bibnamefont
  {Yan}}, \bibinfo {author} {\bibfnamefont {R.}~\bibnamefont {Ravasio}},
  \bibinfo {author} {\bibfnamefont {C.}~\bibnamefont {Brito}},\ and\ \bibinfo
  {author} {\bibfnamefont {M.}~\bibnamefont {Wyart}},\ }\bibfield  {title}
  {\bibinfo {title} {Principles for optimal cooperativity in allosteric
  materials},\ }\href {https://doi.org/10.1016/j.bpj.2018.05.015} {\bibfield
  {journal} {\bibinfo  {journal} {Biophysical Journal}\ }\textbf {\bibinfo
  {volume} {114}},\ \bibinfo {pages} {2787} (\bibinfo {year}
  {2018})}\BibitemShut {NoStop}%
\bibitem [{\citenamefont {Husain}\ and\ \citenamefont
  {Murugan}(2020)}]{husain2020physical}%
  \BibitemOpen
  \bibfield  {author} {\bibinfo {author} {\bibfnamefont {K.}~\bibnamefont
  {Husain}}\ and\ \bibinfo {author} {\bibfnamefont {A.}~\bibnamefont
  {Murugan}},\ }\bibfield  {title} {\bibinfo {title} {Physical constraints on
  epistasis},\ }\href@noop {} {\bibfield  {journal} {\bibinfo  {journal}
  {Molecular Biology and Evolution}\ }\textbf {\bibinfo {volume} {37}},\
  \bibinfo {pages} {2865} (\bibinfo {year} {2020})}\BibitemShut {NoStop}%
\bibitem [{\citenamefont {Anisetti}\ \emph {et~al.}(2023)\citenamefont
  {Anisetti}, \citenamefont {Kandala},\ and\ \citenamefont
  {Schwarz}}]{anisetti2023emergent}%
  \BibitemOpen
  \bibfield  {author} {\bibinfo {author} {\bibfnamefont {V.~R.}\ \bibnamefont
  {Anisetti}}, \bibinfo {author} {\bibfnamefont {A.}~\bibnamefont {Kandala}},\
  and\ \bibinfo {author} {\bibfnamefont {J.}~\bibnamefont {Schwarz}},\
  }\bibfield  {title} {\bibinfo {title} {Emergent learning in physical systems
  as feedback-based aging in a glassy landscape},\ }\href@noop {} {\bibfield
  {journal} {\bibinfo  {journal} {arXiv preprint arXiv:2309.04382}\ } (\bibinfo
  {year} {2023})}\BibitemShut {NoStop}%
\bibitem [{\citenamefont {Rocks}\ \emph {et~al.}(2019)\citenamefont {Rocks},
  \citenamefont {Ronellenfitsch}, \citenamefont {Liu}, \citenamefont {Nagel},\
  and\ \citenamefont {Katifori}}]{rocks2019limits}%
  \BibitemOpen
  \bibfield  {author} {\bibinfo {author} {\bibfnamefont {J.~W.}\ \bibnamefont
  {Rocks}}, \bibinfo {author} {\bibfnamefont {H.}~\bibnamefont
  {Ronellenfitsch}}, \bibinfo {author} {\bibfnamefont {A.~J.}\ \bibnamefont
  {Liu}}, \bibinfo {author} {\bibfnamefont {S.~R.}\ \bibnamefont {Nagel}},\
  and\ \bibinfo {author} {\bibfnamefont {E.}~\bibnamefont {Katifori}},\
  }\bibfield  {title} {\bibinfo {title} {Limits of multifunctionality in
  tunable networks},\ }\href@noop {} {\bibfield  {journal} {\bibinfo  {journal}
  {Proceedings of the National Academy of Sciences}\ }\textbf {\bibinfo
  {volume} {116}},\ \bibinfo {pages} {2506} (\bibinfo {year}
  {2019})}\BibitemShut {NoStop}%
\bibitem [{\citenamefont {Rocks}\ \emph {et~al.}(2020)\citenamefont {Rocks},
  \citenamefont {Liu},\ and\ \citenamefont
  {Katifori}}]{rocks2020StructureFunction}%
  \BibitemOpen
  \bibfield  {author} {\bibinfo {author} {\bibfnamefont {J.~W.}\ \bibnamefont
  {Rocks}}, \bibinfo {author} {\bibfnamefont {A.~J.}\ \bibnamefont {Liu}},\
  and\ \bibinfo {author} {\bibfnamefont {E.}~\bibnamefont {Katifori}},\
  }\bibfield  {title} {\bibinfo {title} {Revealing structure-function
  relationships in functional flow networks via persistent homology},\
  }\bibfield  {journal} {\bibinfo  {journal} {Physical Review Research}\
  }\textbf {\bibinfo {volume} {2}},\ \href
  {https://doi.org/10.1103/physrevresearch.2.033234}
  {10.1103/physrevresearch.2.033234} (\bibinfo {year} {2020})\BibitemShut
  {NoStop}%
\bibitem [{\citenamefont {Rocks}\ \emph {et~al.}(2021)\citenamefont {Rocks},
  \citenamefont {Liu},\ and\ \citenamefont {Katifori}}]{rocks2021hidden}%
  \BibitemOpen
  \bibfield  {author} {\bibinfo {author} {\bibfnamefont {J.~W.}\ \bibnamefont
  {Rocks}}, \bibinfo {author} {\bibfnamefont {A.~J.}\ \bibnamefont {Liu}},\
  and\ \bibinfo {author} {\bibfnamefont {E.}~\bibnamefont {Katifori}},\
  }\bibfield  {title} {\bibinfo {title} {Hidden topological structure of flow
  network functionality},\ }\href@noop {} {\bibfield  {journal} {\bibinfo
  {journal} {Physical Review Letters}\ }\textbf {\bibinfo {volume} {126}},\
  \bibinfo {pages} {028102} (\bibinfo {year} {2021})}\BibitemShut {NoStop}%
\bibitem [{\citenamefont {Rocks}\ \emph {et~al.}(2024)\citenamefont {Rocks},
  \citenamefont {Katifori},\ and\ \citenamefont {Liu}}]{rocks2024topological}%
  \BibitemOpen
  \bibfield  {author} {\bibinfo {author} {\bibfnamefont {J.~W.}\ \bibnamefont
  {Rocks}}, \bibinfo {author} {\bibfnamefont {E.}~\bibnamefont {Katifori}},\
  and\ \bibinfo {author} {\bibfnamefont {A.~J.}\ \bibnamefont {Liu}},\
  }\bibfield  {title} {\bibinfo {title} {Topological characterization of the
  continuum of allosteric response},\ }\href@noop {} {\bibfield  {journal}
  {\bibinfo  {journal} {arXiv preprint arXiv:2401.13861}\ } (\bibinfo {year}
  {2024})}\BibitemShut {NoStop}%
\bibitem [{\citenamefont {Stern}\ \emph {et~al.}(2022)\citenamefont {Stern},
  \citenamefont {Dillavou}, \citenamefont {Miskin}, \citenamefont {Durian},\
  and\ \citenamefont {Liu}}]{stern2021physical}%
  \BibitemOpen
  \bibfield  {author} {\bibinfo {author} {\bibfnamefont {M.}~\bibnamefont
  {Stern}}, \bibinfo {author} {\bibfnamefont {S.}~\bibnamefont {Dillavou}},
  \bibinfo {author} {\bibfnamefont {M.~Z.}\ \bibnamefont {Miskin}}, \bibinfo
  {author} {\bibfnamefont {D.~J.}\ \bibnamefont {Durian}},\ and\ \bibinfo
  {author} {\bibfnamefont {A.~J.}\ \bibnamefont {Liu}},\ }\bibfield  {title}
  {\bibinfo {title} {Physical learning beyond the quasistatic limit},\ }\href
  {https://doi.org/10.1103/PhysRevResearch.4.L022037} {\bibfield  {journal}
  {\bibinfo  {journal} {Phys. Rev. Research}\ }\textbf {\bibinfo {volume}
  {4}},\ \bibinfo {pages} {L022037} (\bibinfo {year} {2022})}\BibitemShut
  {NoStop}%
\bibitem [{\citenamefont {Rouviere}\ \emph {et~al.}(2023)\citenamefont
  {Rouviere}, \citenamefont {Ranganathan},\ and\ \citenamefont
  {Rivoire}}]{rouviere2023emergence}%
  \BibitemOpen
  \bibfield  {author} {\bibinfo {author} {\bibfnamefont {E.}~\bibnamefont
  {Rouviere}}, \bibinfo {author} {\bibfnamefont {R.}~\bibnamefont
  {Ranganathan}},\ and\ \bibinfo {author} {\bibfnamefont {O.}~\bibnamefont
  {Rivoire}},\ }\bibfield  {title} {\bibinfo {title} {Emergence of
  single-versus multi-state allostery},\ }\href@noop {} {\bibfield  {journal}
  {\bibinfo  {journal} {PRX Life}\ }\textbf {\bibinfo {volume} {1}},\ \bibinfo
  {pages} {023004} (\bibinfo {year} {2023})}\BibitemShut {NoStop}%
\bibitem [{\citenamefont {Hexner}\ \emph
  {et~al.}(2020{\natexlab{b}})\citenamefont {Hexner}, \citenamefont {Liu},\
  and\ \citenamefont {Nagel}}]{hexner2019periodic}%
  \BibitemOpen
  \bibfield  {author} {\bibinfo {author} {\bibfnamefont {D.}~\bibnamefont
  {Hexner}}, \bibinfo {author} {\bibfnamefont {A.~J.}\ \bibnamefont {Liu}},\
  and\ \bibinfo {author} {\bibfnamefont {S.~R.}\ \bibnamefont {Nagel}},\
  }\bibfield  {title} {\bibinfo {title} {Periodic training of creeping
  solids},\ }\href@noop {} {\bibfield  {journal} {\bibinfo  {journal}
  {Proceedings of the National Academy of Sciences}\ }\textbf {\bibinfo
  {volume} {117}},\ \bibinfo {pages} {31690} (\bibinfo {year}
  {2020}{\natexlab{b}})}\BibitemShut {NoStop}%
\bibitem [{\citenamefont {Halabi}\ \emph {et~al.}(2009)\citenamefont {Halabi},
  \citenamefont {Rivoire}, \citenamefont {Leibler},\ and\ \citenamefont
  {Ranganathan}}]{halabi2024protein}%
  \BibitemOpen
  \bibfield  {author} {\bibinfo {author} {\bibfnamefont {N.}~\bibnamefont
  {Halabi}}, \bibinfo {author} {\bibfnamefont {O.}~\bibnamefont {Rivoire}},
  \bibinfo {author} {\bibfnamefont {S.}~\bibnamefont {Leibler}},\ and\ \bibinfo
  {author} {\bibfnamefont {R.}~\bibnamefont {Ranganathan}},\ }\bibfield
  {title} {\bibinfo {title} {Protein sectors: Evolutionary units of
  three-dimensional structure},\ }\href@noop {} {\bibfield  {journal} {\bibinfo
   {journal} {Cell}\ ,\ \bibinfo {pages} {774}} (\bibinfo {year}
  {2009})}\BibitemShut {NoStop}%
\bibitem [{\citenamefont {Alim}\ \emph {et~al.}(2017)\citenamefont {Alim},
  \citenamefont {Andrew}, \citenamefont {Pringle},\ and\ \citenamefont
  {Brenner}}]{alim2017mechanism}%
  \BibitemOpen
  \bibfield  {author} {\bibinfo {author} {\bibfnamefont {K.}~\bibnamefont
  {Alim}}, \bibinfo {author} {\bibfnamefont {N.}~\bibnamefont {Andrew}},
  \bibinfo {author} {\bibfnamefont {A.}~\bibnamefont {Pringle}},\ and\ \bibinfo
  {author} {\bibfnamefont {M.~P.}\ \bibnamefont {Brenner}},\ }\bibfield
  {title} {\bibinfo {title} {Mechanism of signal propagation in physarum
  polycephalum},\ }\href@noop {} {\bibfield  {journal} {\bibinfo  {journal}
  {Proceedings of the National Academy of Sciences}\ }\textbf {\bibinfo
  {volume} {114}},\ \bibinfo {pages} {5136} (\bibinfo {year}
  {2017})}\BibitemShut {NoStop}%
\bibitem [{\citenamefont {Anisetti}\ \emph {et~al.}(2022)\citenamefont
  {Anisetti}, \citenamefont {Kandala}, \citenamefont {Scellier},\ and\
  \citenamefont {Schwarz}}]{anisetti2022frequency}%
  \BibitemOpen
  \bibfield  {author} {\bibinfo {author} {\bibfnamefont {V.~R.}\ \bibnamefont
  {Anisetti}}, \bibinfo {author} {\bibfnamefont {A.}~\bibnamefont {Kandala}},
  \bibinfo {author} {\bibfnamefont {B.}~\bibnamefont {Scellier}},\ and\
  \bibinfo {author} {\bibfnamefont {J.}~\bibnamefont {Schwarz}},\ }\bibfield
  {title} {\bibinfo {title} {Frequency propagation: Multi-mechanism learning in
  nonlinear physical networks},\ }\href@noop {} {\bibfield  {journal} {\bibinfo
   {journal} {arXiv preprint arXiv:2208.08862}\ } (\bibinfo {year}
  {2022})}\BibitemShut {NoStop}%
\bibitem [{\citenamefont {Hornik}\ \emph {et~al.}(1989)\citenamefont {Hornik},
  \citenamefont {Stinchcombe},\ and\ \citenamefont
  {White}}]{hornik1989multilayer}%
  \BibitemOpen
  \bibfield  {author} {\bibinfo {author} {\bibfnamefont {K.}~\bibnamefont
  {Hornik}}, \bibinfo {author} {\bibfnamefont {M.}~\bibnamefont
  {Stinchcombe}},\ and\ \bibinfo {author} {\bibfnamefont {H.}~\bibnamefont
  {White}},\ }\bibfield  {title} {\bibinfo {title} {Multilayer feedforward
  networks are universal approximators},\ }\href@noop {} {\bibfield  {journal}
  {\bibinfo  {journal} {Neural networks}\ }\textbf {\bibinfo {volume} {2}},\
  \bibinfo {pages} {359} (\bibinfo {year} {1989})}\BibitemShut {NoStop}%
\bibitem [{\citenamefont {Stern}\ \emph
  {et~al.}(2024{\natexlab{c}})\citenamefont {Stern}, \citenamefont {Dillavou},
  \citenamefont {Jayaraman}, \citenamefont {Durian},\ and\ \citenamefont
  {Liu}}]{stern2024training}%
  \BibitemOpen
  \bibfield  {author} {\bibinfo {author} {\bibfnamefont {M.}~\bibnamefont
  {Stern}}, \bibinfo {author} {\bibfnamefont {S.}~\bibnamefont {Dillavou}},
  \bibinfo {author} {\bibfnamefont {D.}~\bibnamefont {Jayaraman}}, \bibinfo
  {author} {\bibfnamefont {D.~J.}\ \bibnamefont {Durian}},\ and\ \bibinfo
  {author} {\bibfnamefont {A.~J.}\ \bibnamefont {Liu}},\ }\bibfield  {title}
  {\bibinfo {title} {Training self-learning circuits for power-efficient
  solutions},\ }\href@noop {} {\bibfield  {journal} {\bibinfo  {journal} {APL
  Machine Learning}\ }\textbf {\bibinfo {volume} {2}} (\bibinfo {year}
  {2024}{\natexlab{c}})}\BibitemShut {NoStop}%
\bibitem [{\citenamefont {Liu}\ and\ \citenamefont
  {Nagel}(2010)}]{liu2010jamming}%
  \BibitemOpen
  \bibfield  {author} {\bibinfo {author} {\bibfnamefont {A.~J.}\ \bibnamefont
  {Liu}}\ and\ \bibinfo {author} {\bibfnamefont {S.~R.}\ \bibnamefont
  {Nagel}},\ }\bibfield  {title} {\bibinfo {title} {The jamming transition and
  the marginally jammed solid},\ }\href@noop {} {\bibfield  {journal} {\bibinfo
   {journal} {Annu. Rev. Condens. Matter Phys.}\ }\textbf {\bibinfo {volume}
  {1}},\ \bibinfo {pages} {347} (\bibinfo {year} {2010})}\BibitemShut {NoStop}%
\end{thebibliography}%

\end{document}